\documentclass[aps,prb,twocolumn,9pt,superscriptaddress,citeautoscript]{revtex4-2}

\pdfoutput=1
\usepackage{graphicx}
\usepackage{amsmath, amsfonts}
\usepackage{amssymb}
\usepackage[colorlinks=true,linkcolor=blue,citecolor=blue]{hyperref}
\usepackage[utf8]{inputenc}
\usepackage{color}

\begin{document}

\title{Quantum transport properties of tantalum-oxide resistive switching filaments}

\author{Tímea Nóra Török}
\affiliation{Department of Physics, Institute of Physics, Budapest University of Technology and Economics, M\H{u}egyetem rkp. 3., H-1111 Budapest, Hungary.}
\affiliation{Institute of Technical Physics and Materials Science, Centre for Energy Research, Konkoly-Thege M. \'{u}t 29-33, 1121 Budapest, Hungary.}

\author{Péter Makk}
\affiliation{Department of Physics, Institute of Physics, Budapest University of Technology and Economics, M\H{u}egyetem rkp. 3., H-1111 Budapest, Hungary.}
\affiliation{MTA-BME Correlated van der Waals Structures Momentum Research Group, M\H{u}egyetem rkp. 3., H-1111 Budapest, Hungary.}

\author{Zoltán Balogh}
\affiliation{Department of Physics, Institute of Physics, Budapest University of Technology and Economics, M\H{u}egyetem rkp. 3., H-1111 Budapest, Hungary.}
\affiliation{ELKH-BME Condensed Matter Research Group, M\H{u}egyetem rkp. 3., H-1111 Budapest, Hungary.}

\author{Miklós Csontos}
\affiliation{Institute of Electromagnetic Fields, ETH Zurich, Gloriastrasse 35, 8092 Zurich, Switzerland.}

\author{András Halbritter$^{\ast}$}
\affiliation{Department of Physics, Institute of Physics, Budapest University of Technology and Economics, M\H{u}egyetem rkp. 3., H-1111 Budapest, Hungary.}
\affiliation{ELKH-BME Condensed Matter Research Group, M\H{u}egyetem rkp. 3., H-1111 Budapest, Hungary.}

\begin{abstract}
{Filamentary resistive switching devices are not only considered as promising building blocks for brain-inspired computing architectures, but they also realize an unprecedented operation regime, where the active device volume reaches truly atomic dimensions. Such atomic-sized resistive switching filaments represent the quantum transport regime, where the transmission eigenvalues of the conductance channels are considered as a specific device fingerprint. Here, we gain insight into the quantum transmission properties of close-to-atomic-sized resistive switching filaments formed across an insulating Ta$_2$O$_5$ layer through superconducting subgap spectroscopy. This method reveals the transmission density function of the open conduction channels contributing to the device conductance. Our analysis confirms the formation of truly atomic-sized filaments composed of 3\,--\,8 Ta atoms at their narrowest cross-section. We find that this diameter remains unchanged upon resistive switching. Instead, the switching is governed by the redistribution of oxygen vacancies within the filamentary volume. The set/reset process results in the reduction/formation of an extended barrier at the bottleneck of the filament which enhances/reduces the transmission of the highly open conduction channels.}
\end{abstract}

\date{\today}
\maketitle


\section{Introduction}

Resistive switching (RS) memory devices (a.k.a. ``memristors'') are identified as a major hardware platform enabling the power-efficient implementation of artificial intelligence. Pioneering applications employ large memristor crossbar arrays in artificial neural networks, which implement machine learning approaches at the hardware level at unprecedented speed and energy efficiency.~\cite{Yang2013,Zidan2018,Ambrogio2018,Xia2019} The neuromorphic functionalities of single memristors arise from the dynamical properties of a previously unavailable operation regime, where the active device volume can reach truly atomic dimensions.~\cite{Torok2020,Milano2022,Aono2010,Cheng2019,res-switch-book} This ultra-small active region is manifested by close-to-atomic-sized metallic filaments which can be formed or ruptured by appropriate voltage signals. Such ``artificial synapses'' are utilized to encode synaptic weights through their analog tunable filamentary conductance.

Beyond the synaptic functionalities of RS devices, the \emph{metallic} nature of the filamentary volume yields a completely different \emph{quantum transport regime} compared to semiconductor heterostructures: (i) The large electron density ($\approx 1$ electron/lattice atom) results in orders of magnitude larger Fermi energies ($E_F^\mathrm{metal}\approx 3-10\,$eV~\cite{Kittel2005}) than in semiconducting systems. Therefore, quantum transport is realized at room temperature due to the $E_F \gg k_B T$ condition. (ii) Unlike in semiconductor heterostructures, the metallic Fermi wavelength ($\lambda_F^\mathrm{metal}\approx 0.3-0.6\,$nm) falls in the regime of the lattice constant. Accordingly, the electrons experience a potential landscape which is rough on the scale of their wavelength. This results in far more complex quantum transport properties than universal conductance quantization in semiconductor quantum point contacts,~\cite{vanWees1988} where a smooth, transparent quantum channel is realized. This complex quantum transport behavior has been widely investigated in pure metallic atomic-sized nanowires,~\cite{Ruitenbeek-PhysRep-quantum,Scheer-Cuevas, Scheer1998} but has remained poorly explored in filamentary RS devices.

In this paper we investigate the quantum transport properties of RS filaments formed across a Ta$_2$O$_5$ thin film due to the voltage induced redistribution of oxygen vacancies.\cite{Park2013,Marchewka2016} Ta$_2$O$_5$ is a prominent representative of transition metal oxides which are widely considered as a competitive material platform for neuromorphic hardware architectures.~\cite{Yang2010,Torrezan2011,Strachan2011,Lee2011,Graves2017,Rao2022,Csontos2022} In order to track the atomistic changes in the filament structure taking place upon resistive switching, we apply superconducting subgap spectroscopy.\cite{Scheer1997,Scheer1998,Ludoph2000,Ruitenbeek-PhysRep-quantum,Makk2008,Torok2020} This method enables the analysis of the distribution of the $\tau_i$ quantum transmission eigenvalues of the conduction channels contributing to the conductance of the RS filament. The total filamentary conductance is obtained via the Landauer formula~\cite{Landauer1970} as $G=G_0\cdot\sum_{i=1}^{M}\tau_i$, where $G_0=2e^2/h$ is the quantum conductance unit and $M$ is the number of open conductance channels. We focus on RS filaments, where $M$ is sufficiently large so that the transmission eigenvalues can be described by a probability density function $\rho^{\textrm{Ta}_2\textrm{O}_5}(\tau)$. We determine $\rho^{\textrm{Ta}_2\textrm{O}_5}(\tau)$ in the high (HCS) and low (LCS) conductance states and compare to that of pure Ta nanowires, $\rho^\textrm{Ta}(\tau)$. Whereas a close-to-atomic-scale metallic nanowire is responsible for the conduction in both systems, the formation and the environment of this metallic wire is fundamentally different. The reference Ta nanowires represent a pure metallic system which is thinned by mechanical elongation in vacuum conditions. In contrast, the oxide-based RS filaments are embedded in an insulating, amorphous Ta$_2$O$_5$ matrix. The evolution of the RS junction relies on the voltage-induced displacement of oxygen vacancies, and thereby the formation/destruction of oxygen-deficient conducting filamentary regions. Consequently, our analysis reveals a fundamentally different conductance channel distribution for these two systems. In pure Ta nanowires  $\rho^\textrm{Ta}(\tau)$ resembles the universal probability density function $\rho^\mathrm{RMT}(\tau)\sim\tau^{-1}(1-\tau)^{-1/2}$ in a broad diameter range. The latter formula was derived for mesoscopic diffusive wires by random matrix theory.\cite{Beenakker1997} In this case point-like scattering centers are considered which are, however, bypassed by certain conductance channels yielding almost perfect transmission ($\tau_i\approx 1$) for a subset of the channels. In contrast, in RS filaments the formation of an extended barrier across the narrowest section of the filament is revealed upon the reset transition while the filament diameter is hardly affected. This barrier is attributed to the accumulation of oxygen ions. It reduces the transmission of the highly opened conduction channels, yielding a completely different transmission distribution than that of pure Ta nanowires.

\section{Results and Discussion}

\begin{figure}[t!]
	\centering
	\includegraphics[width=0.48\textwidth]{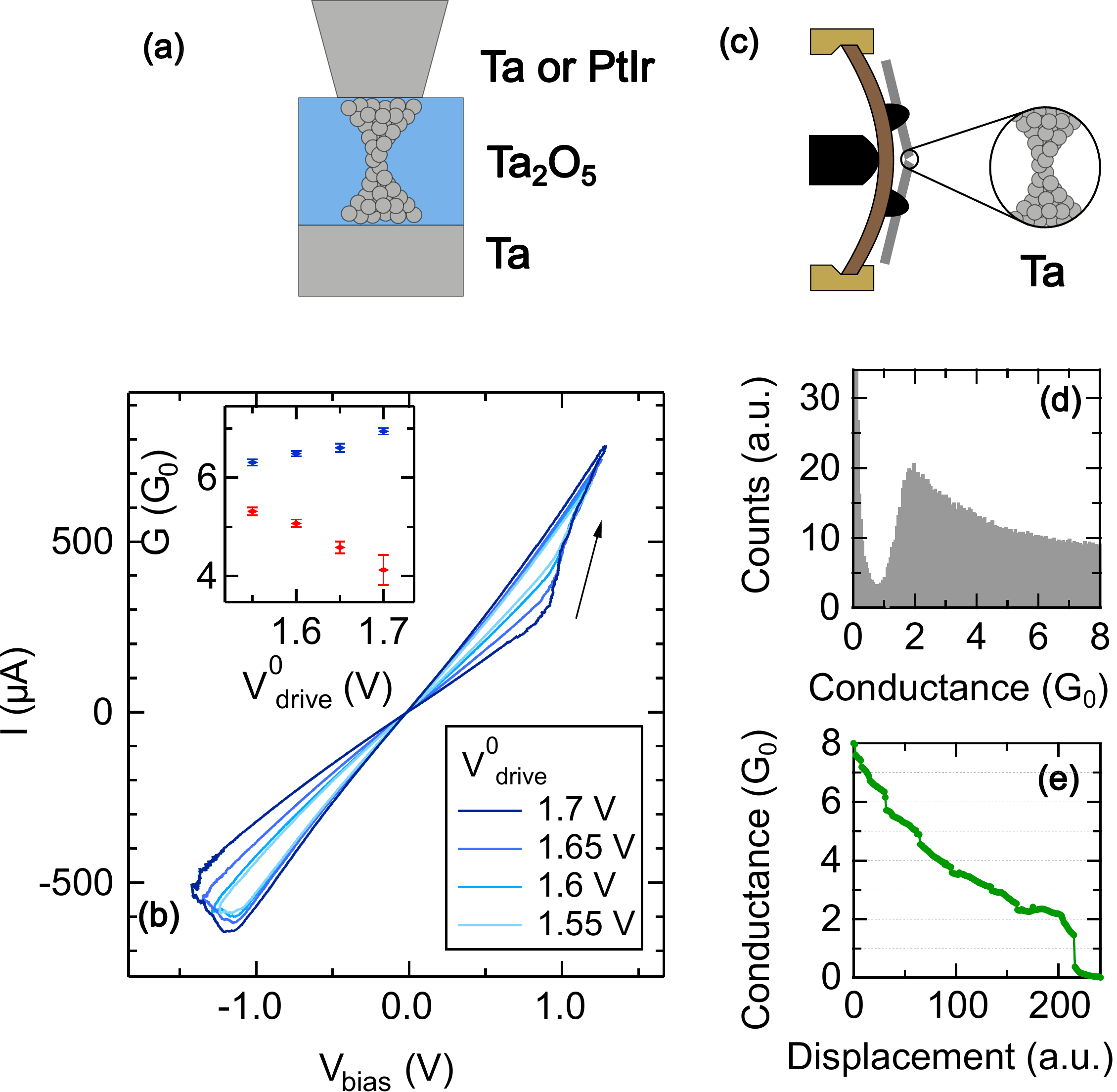}
	\caption{\it \textbf{Characterization of RS filaments in Ta$_2$O$_5$ and pure Ta atomic wires.} (a) Illustration of a Ta$_2$O$_5$ RS device realized in an STM point-contact arrangement. (b) Representative room temperature RS current-voltage characteristics using a conventional PtIr tip. The hysteretic switching trace opens up as the amplitude of the triangular driving voltage signal is increased (see the color-scale). The arrow indicates the direction of the hysteresis. The inset shows the low-bias conductance of the HCS (blue) and LCS (red) at each voltage amplitude demonstrating multilevel programming (average and standard deviation of 10 consecutive switching cycles are shown for each voltage). (c) Illustration of pure Ta atomic wires established by mechanical rupture using the three point bending arrangement in a MCBJ setup. (d) A representative conductance histogram of pure Ta nanowires obtained by collecting thousands of conductance versus electrode separation traces during the repeated opening and closing of the junction. The cryogenic vacuum at liquid helium temperature grants the extreme cleanliness during the in-situ rupture. (e) An exemplary conductance versus electrode separation trace.}
	\label{fig1}
\end{figure}

\textit{\textbf{Experimental approach.}} For our comparative analysis we established atomic-sized RS filaments in Ta$_2$O$_5$ and pure Ta atomic wires by an STM (scanning tunneling microscopy) point contact setup and a mechanically controllable break junction (MCBJ) setup, respectively (see Fig.~\ref{fig1}a,c). In order to apply the method of superconducting subgap spectroscopy, both systems were cooled to a base temperature of $T\approx1.3\,$K, i.e. well below the superconducting phase transition of tantalum ($T_C=4.48\,$K). The current-voltage [$I(V)$] characteristics of a statistical ensemble of junctions were recorded in the superconducting state and fitted by simulated $I(V)$ curves relying on model transmission distributions. In the following, we first describe our experimental approach. Next, we discuss the applied model transmission distributions and the fitting procedure. Finally, we describe and discuss the results of our analysis.

RS was studied in an STM point-contact setup, where PtIr or Ta tips were approached and touched to Ta$_2$O$_5$/Ta thin film samples in a controlled manner. The PtIr tips were applied in routine room temperature characterization measurements, whereas Ta tips were used in the low temperature measurements to have superconducting electrodes at both sides of the RS filament. The resistive switching layer was prepared by sputtering $\sim$30~nm thick stoichiometric Ta$_2$O$_5$ on top of a $30\,$nm thick Ta layer (see Fig.~\ref{fig1}a). The approach of the STM tip was stopped at a setpoint resistance of $1\,$G$\Omega$. Afterwards, the $I(V)$ characteristics with typical voltage driving amplitudes of $V^0_\mathrm{drive}\approx1-2\,$V readily exhibited resistive switching without a dedicated high-voltage electroforming process. The set process was terminated by a series resistor of $R_S=300-3600\,\Omega$. Accordingly, the bias voltage on the sample is calculated as $V_\textrm{bias}=V_\textrm{drive}-I\cdot R_S$. The details of the switching were fine-tunable either by the gentle manipulation of the STM tip or the variation of the driving voltage amplitude. Room temperature resistive switching characteristics at fixed tip position and varying driving amplitude are exemplified in Fig.~\ref{fig1}b. Positive bias refers to a higher potential acting on the thin film with respect to the STM tip. The inset demonstrates the low-bias conductance values in the LCS and HCS of the hysteretic $I(V)$ characteristics demonstrating fine analog tunability in the $4~G_0-7~G_0$ range. Our measurements on several resistive switching junctions focused on the similar, $\approx1~G_0-15~G_0$ conductance regime, where the quantum transport properties can be compared to those of atomic-sized pure tantalum nanowires. Measurements utilizing Ta STM tips exhibited similar characteristics both at room temperature (see Fig.~S1 in the SI) and at cryogenic temperatures (see Fig.~\ref{fig3} and its discussion). Note, that the above conductance regime is also ideal for resistive switching synapses in artificial neural networks, where high-accuracy hardware-based vector-matrix multiplication operation requires linear $I(V)$ characteristics below the resistive switching threshold voltage.~\cite{Yang2013,Zidan2018,Li2018} 

As a reference system, we have established pure Ta atomic-sized wires in a MCBJ arrangement. In this setup a notched Ta wire with $0.1\,$mm diameter is broken in a three point bending configuration (see Fig.~\ref{fig1}c) using the combination of a coarse stepper motor actuation and a fine piezo control. To characterize the such established Ta nanowires, thousands of conductance vs. electrode separation traces (see an example trace in Fig.~\ref{fig1}e) are recorded along the repeated opening and closing of the junction. The conductance data of these traces are used to construct a conductance histogram (Fig.~\ref{fig1}d). Note, that both the voltage controlled RS filaments (see Fig.~\ref{fig1}b and a further example covering the $\approx1~G_0-3~G_0$ regime in Fig.~S2 of the SI) and the mechanically actuated pure Ta nanowires exhibit a fine analog tunability of the conductance in the entire investigated conductance range lacking any conductance quantization features.

The $I(V)$ characteristics of the superconductor (S) - nanoconstriction (c) - superconductor (S) junctions were acquired according to the same protocol as in our previous work on the subgap analysis of niobium-oxide resistive switching filaments.\cite{Torok2020} The latter work focused on the investigation of RS filaments with truly single-atom cross-section dominated by a single conductance channel. Here, we investigate somewhat larger RS filaments with significantly more open conductance channels and analyze the probability density of the transmission eigenvalues, $\rho(\tau)$. We exploit that in an ScS junction the current contributions of multiple Andreev reflection processes strongly depend on the corresponding transmission eigenvalues. Consequently, the superconducting $I(V)$ characteristics provide a sensitive diagnostic tool to investigate $\rho(\tau)$ (see the computed ScS $I(V)$ curves in Fig. S3 of the SI). 

\begin{figure}[t!]
	\centering
	\includegraphics[width=0.48\textwidth]{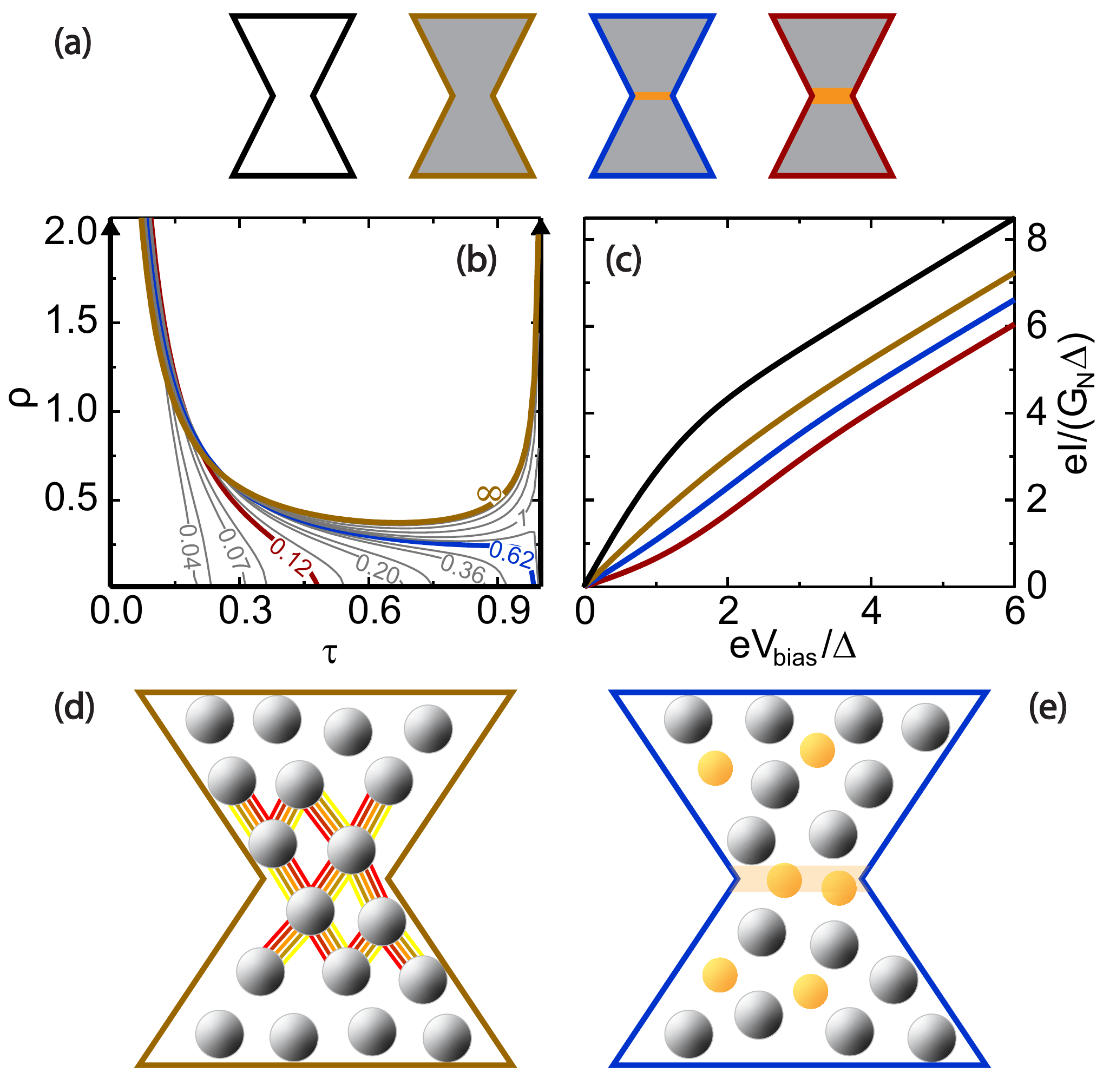}
	\caption{\it \textbf{Model approach for the analysis of the quantum conductance channel distributions.} Panel (a) illustrates possible model geometries: (i) a ballistic quantum wire (black frame), (ii) a diffusive wire without any extended barrier (brown frame), (iii) a diffusive wire with weaker/stronger extended barrier at the bottleneck (blue/red frame). The corresponding model transmission density functions are illustrated in panel (b): the black Dirac delta peaks illustrate the case of quantized conductance channels in a ballistic quantum wire. The brown curve shows the $\rho^\mathrm{RMT}(\tau)$ density function for a diffusive wire. The gray curves including the blue/red example curves represent the density functions in case of an extended barrier with variable strength, as labeled by the $\alpha$ parameters. Panel (c) shows the calculated ScS $I(V)$ curves at $G=10\,$G$_0$ for $\rho^\mathrm{RMT}(\tau)$ (brown), for a filament with a weaker/stronger extended barrier (blue/red curves corresponding to the blue/red density functions highlighted in panel (b)), and for a highly transparent junction with perfectly open conductance channels (black). In the latter case an ideally infinite zero bias slope is expected. The apparent finite slope is a result of the finite energy resolution included in our model. Panels (d) and (e) respectively illustrate a pure Ta  atomic wire and an oxide-based resistive switching filament. In (d) the five colored lines connecting two neighbor atoms illustrate the typical transport through 5 non-vanishing inter-atomic conductance channels in a $d$-valent metal. The transmission eigenvalues of such 5 inter-atomic channels strongly scatter between zero and unity transmission, giving rise to a diffusive-like transport. In the oxide-based RS filaments (e) the relocation of oxygen atoms play a key role in the switching process, also influencing the transport properties. The eventual formation of an extended barrier (as illustrated by the orange region) is attributed to the voltage-controlled accumulation of oxygen.} \label{fig2}
\end{figure}

\textit{\textbf{Model analysis of the transmission distributions.}} Our modeling approach for the fitting of the superconducting $I(V)$ characteristics is illustrated in Fig.~\ref{fig2}. As a reference, we rely on the $\rho^\mathrm{RMT}(\tau)\sim\tau^{-1}(1-\tau)^{-1/2}$ universal probability density function of the transmission eigenvalues (see brown curve in Fig.~\ref{fig2}b). This density function was derived for phase coherent mesoscopic diffusive wires with a large number of open conductance channels ($M\gg 1$). The corresponding model system is illustrated on the second panel of Fig.~\ref{fig2}a (brown frame), where the gray background symbolizes the diffusive nature of the transport. The RMT density function is bimodal: $\rho^\mathrm{RMT}(\tau)$ is peaked at both the low-transmission ($\tau \ll 1$) and high-transmission ($\tau \approx 1$) ends while the intermediate region also exhibits a finite weight. The presence of completely open conductance channels, which was indeed confirmed by shot noise measurements,\cite{Henny1999} seems counter-intuitive due to the diffusive nature of the transport, i.e., the short mean free path compared to the system size. However, it can be shown, that point-like defects can be bypassed by certain conductance channels, yielding almost perfect channel transmission.\cite{Beenakker1997} It was recognized, that this universal density function also well approximates the transmission distribution in Pb atomic-sized wires,\cite{Rubio-Bollinger-subgap} where the disordered nature is attributed to the interplay of non-perfect crystalline ordering, non-perfect transmission especially through the highly oriented $p$ orbitals, and surface scattering. {We anticipate similar behavior in transition metals, like the investigated pure Ta nanowires, where the electron transmission through two neighbor atoms is shared among $\approx5$ conductance channels, as illustrated by the 5 colored lines between two neighbor atoms in Fig.~\ref{fig2}d. This expectation relies on the tight-binding model of a single-atom nanowire\cite{Cuevas_PRL_1998,Scheer-Cuevas} identifying the number of channels with the number of valence orbitals at the central atom (i.e. 1 $s$-type and 5 $d$-type channels in a transition metal), and predicting the closing of one channel due to symmetry reasons. This specific behavior was experimentally verified in Nb and Ta junctions, which have very similar electron structure and conduction properties.\cite{Scheer1998,Makk2008} Note, that the above considerations imply $\approx 10-15$ non-vanishing conductance channels for a filament with only $2-3$ atoms in its narrowest cross-section. Such a large channel number justifies the applicability of a transmission density function.} 

Our goal is to experimentally verify the universal transmission density function in pure Ta atomic wires, and to identify and understand possible deviations from this in oxide-based RS filaments. Since the (re-)distribution of oxygen impurities (see Fig.~\ref{fig2}e) plays important roles both in the transport properties and in the resistive switching process, such a quantitative comparison will enable to draw conclusions about the atomic-scale evolution of the filament upon resistive switching. To this end, we consider possible model transmission density functions both with \emph{more transmissive} and \emph{less transmissive} characteristics (see Figs.~\ref{fig2}a,b). In the former case the most extreme situation is illustrated: a fully ballistic junction (left panel in Fig.~\ref{fig2}a, where the white background symbolizes the ballistic nature). This scenario corresponds to the case of universal conductance quantization, where all open conductance channels exhibit perfect transmission, as demonstrated by the black density function with Dirac-delta peaks at $\tau=0$ and $\tau=1$ in Fig.~\ref{fig2}b. The situation where the overall transmission is lowered compared to the RMT case can be considered according to the model of Nazarov and coworkers,\cite{Nazarov1994} where an extended barrier is formed at the bottleneck of the filament. This results in an enhanced back-scattering in all the channels, yielding a suppression of the peak at $\tau \approx 1$ in the $\rho^\mathrm{RMT}(\tau)$ transmission distribution. This model relies on the series connection of a diffusive wire with $R_W$ resistance and an extended barrier with $R_B$ resistance. The correspondingly derived transmission density function depends on the $\alpha=R_W/R_B$ parameter, as demonstrated by the gray curves (including the example curves highlighted in red and blue) in Fig.~\ref{fig2}b. The corresponding model systems are illustrated on the right two panels of Fig.~\ref{fig2}a, where the orange region symbolizes the extended barrier. Note, that the $\alpha\rightarrow\infty$ limit recovers the universal $\rho^\mathrm{RMT}(\tau)$ transmission density function. 

In order to experimentally determine the relevance of the above described scenarios in our devices, we compare the measured $I(V)$ traces to simulated model $I(V)$ traces corresponding to a given transmission density function as follows. By choosing a certain model transmission density function $\rho(\tau)$ normalized to unity ($\int\rho(\tau)\mathrm{d}\tau$=1), the average transmission probability of the open channels is determined as $\overline{\tau_\rho}=\int \tau \rho(\tau) \mathrm{d}\tau$. Relying on this average transmission, the number of open channels at a certain $G$ conductance is estimated as $M_{\rho,G}=\mathrm{round}\left((G/G_0)/\overline{\tau_\rho}\right)$. The transmission eigenvalues of an atomic filament/wire with $G$ conductance are simulated by generating $M_{\rho,G}$ random numbers according to the chosen $\rho(\tau)$ probability density function. Having these random transmission eigenvalues at hand, the $I(V)$ curve of the hypothetical ScS junction characterized by the particular $\tau_i$ values is simulated according to the multiple Andreev reflection theory.\cite{Cuevas1996,Averin1995,Rubio-Bollinger-subgap} For the sake of a meaningful comparison to experimental $I(V)$ data, an ensemble averaging is performed by repeating the above approach for a large number of independent sets of random transmission eigenvalues. Thereby an average ScS $I(V)$ characteristic can be modeled.

Additionally, our analysis also takes into account the finite energy resolution of the measurements arising from the proximity effect\cite{Torok2020} by smoothing the modeled $I(V)$ curves with a Gaussian kernel according to the actual energy resolution (see further details in Fig. S3 of the SI).  Furthermore, our numerical analysis applies a bottom cutoff at $\tau=0.03$, i.e., channels with smaller transmission are not considered as open. With this physically reasonable choice $\rho^\mathrm{RMT}(\tau)$ yields $\overline{\tau_\rho}=0.39$, which reproduces the expected $M=5$ open channels for a single-atom junction corresponding to the conductance histogram peak in Fig.~\ref{fig1}d. The value of $\overline{\tau_\rho}$ and $M$ obviously depend on the choice of the cutoff, but the results of the subgap analysis are insensitive to the precise cutoff value (see Fig.~S4 of the SI).

Figure~\ref{fig2}c shows the such simulated $I(V)$ characteristics for hypothetical ScS junctions with $G=10\,G_0$ conductance and various transmission density functions. The brown curve illustrates the case of the universal $\rho^\mathrm{RMT}(\tau)$ density function corresponding to the brown curve in Fig.~\ref{fig2}b. The blue and red curves correspond to junctions including extended barriers with $\alpha=0.62$ and $\alpha=0.12$, respectively (see the blue and red $\rho(\tau)$ curves in Fig.~\ref{fig2}b). As a reference, the black curve corresponds to a junction satisfying universal conductance quantization, where all the open conductance channels have $\tau_i=1$ transmission. The ScS $I(V)$ trace shows markedly different behavior for the above, fundamentally different transmission densities. By normalizing the $I(V)$ curves with the $\Delta$ superconducting gap and the $G_N$ normal state conductance (see the axes in Fig.~\ref{fig2}c) a unity slope line is obtained in the normal state for any filament type. Superconductivity, however, enhances the contribution of open channels ($\tau_i\approx 1$) and suppresses the contribution of weakly transmitting channels ($\tau_i\ll 1$). Accordingly, a junction with highly / poorly transmitting channels (black / red) exhibits a considerably larger / smaller slope than unity around zero bias ($eV\ll 2\Delta$). As another dominant difference, the highly transmitting junction also exhibits a considerable excess current, i.e., the high-bias linear section of the $I(V)$ curve extrapolates to a significant positive current at zero bias.\cite{Senkpiel2022} The universal $\rho^\mathrm{RMT}(\tau)$ density yields an $I(V)$ curve in-between these two extremities, with moderately larger than unity slope at zero bias, and a moderate excess current at high bias.

\textit{\textbf{Results of the subgap spectroscopy measurements.}} Figure~\ref{fig3} displays the subgap analysis of the transmission density functions based on the ScS $I(V)$ curves measured on RS filaments (Fig.~\ref{fig3}c) and pure Ta atomic wires (Fig.~\ref{fig3}d). As a reference, first the data of the latter system is analyzed. The dotted line shows the measured ScS $I(V)$ curve, which very much resembles the simulated $I(V)$ curve according to the universal $\rho^\mathrm{RMT}(\tau)$ density function (brown dashed line). A slightly better agreement is obtained according to Nazarov's theory,\cite{Nazarov1994} using the best fitting $\alpha=1.6$  parameter (green dashed line). The corresponding transmission density function (green line in Fig.~\ref{fig3}b) is indeed positioned very close to the $\rho^\mathrm{RMT}(\tau)$ universal density function (brown curve in Fig.~\ref{fig3}b). This analysis supports our preconception, that Ta atomic wires are well described by the universal transmission density function. For this particular junction the $G=8.33\,$G$_0$ conductance and the universal transmission density function yields $M\approx 21$ open channels, indicating a junction with $\approx 4$ atoms at the narrowest cross-section according to the $5$ channels/atom estimate.

\begin{figure}[t!]
	\centering
	\includegraphics[width=0.475\textwidth]{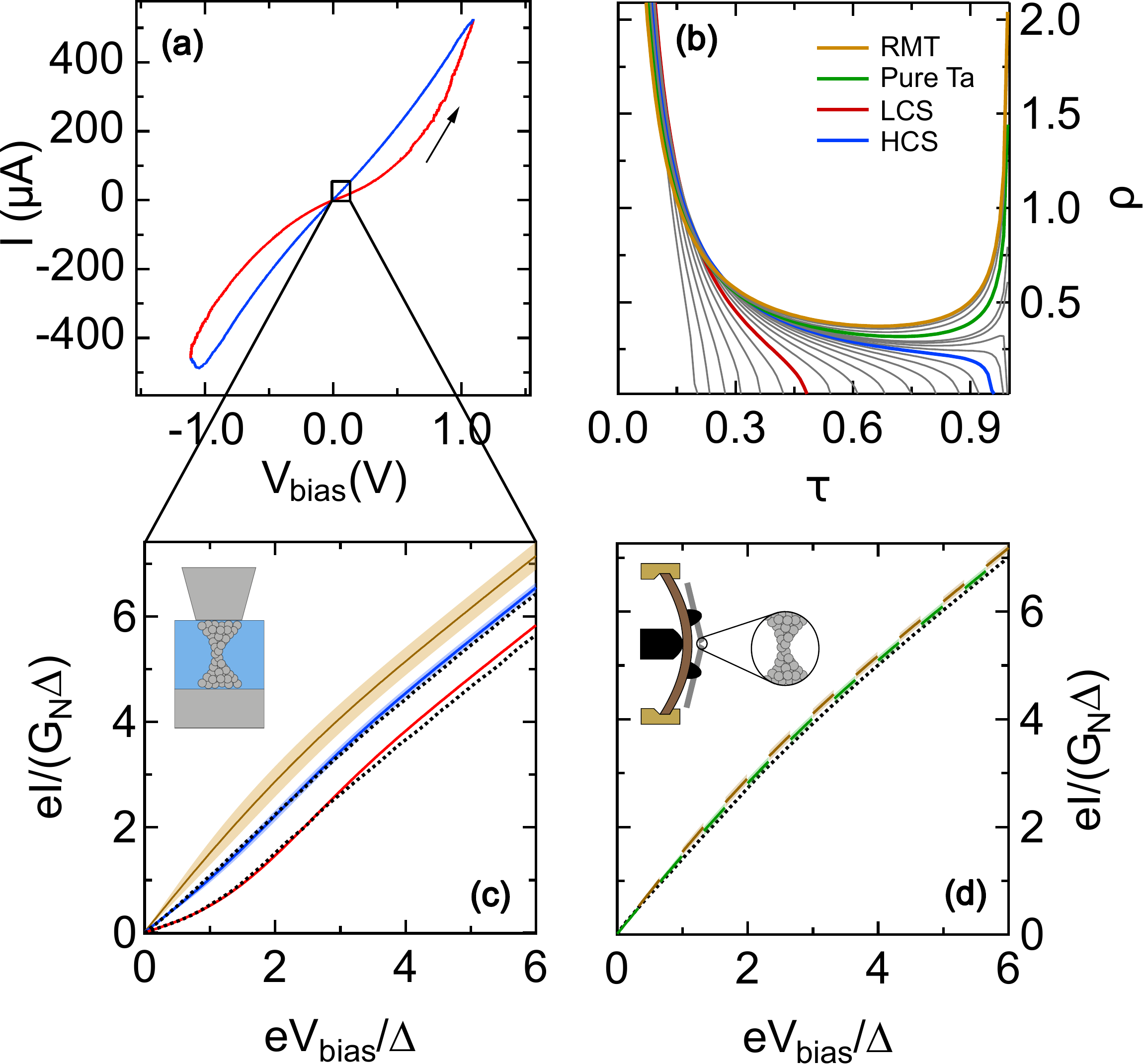}
	\caption{\it \textbf{Subgap analysis of the conductance channels in a representative RS filament and a pure Ta atomic wire.} (a) Typical resistive switching curve of Ta(tip)/Ta$_2$O$_5$/Ta STM junctions at cryogenic temperature. The arrow indicates the direction of the hysteresis. (b) The red, blue and green lines demonstrate the best fitting transmission densities for the LCS and HCS of the RS filament in panel (c) and for the pure Ta atomic wire in panel (d), respectively. As a reference, the light brown curve shows $\rho^\mathrm{RMT}(\tau)$. (c) ScS $I(V)$ curves for there HCS (top dotted line, $G=4.03\,$G$_0$) and the LCS (bottom dotted line, $G=2.02\,$G$_0$) of the RS $I(V)$ curve in panel (a). The blue and red curves represent the best fitting model $I(V)$ curves, wheres the light brown curve shows the model $I(V)$ curve for $\rho^\mathrm{RMT}(\tau)$ as a reference. For all the colored curves the light colored background region represents the standard deviation of the $I(V)$ curves for the same density function, but various random transmission sets. (d) The measured ScS $I(V)$ curve (dotted line, $G=8.33\,$G$_0$) and the best fitting model $I(V)$ curve (green dashed line) for the reference pure Ta atomic wire. The measured $I(V)$ curve very much resembles the model $I(V)$ curve for $\rho^\mathrm{RMT}(\tau)$ (light brown dashed line).}
	\label{fig3}
\end{figure}

Figure~\ref{fig3}a exemplifies a cryogenic temperature ($T\approx1.3\,$K) resistive switching curve, where the HCS and LCS are shown in blue and red colors, respectively. Resistive switching occurs on the voltage-scale of $1\,$V, whereas the superconducting features are detected on a much smaller voltage scale, $V\approx 2\Delta/e\approx 1.33\,$mV.~\cite{Levy1963} The top and bottom dotted lines in Fig.~\ref{fig3}c show the measured ScS $I(V)$ curves in the latter voltage range for the HCS and the LCS, respectively. The blue and red lines represent the best fitting model $I(V)$ curves according to Nazarov's theory, using $\alpha=0.51$ (HCS, blue curve) and $\alpha=0.12$ (LCS, red curve). The corresponding transmission density functions are shown in Fig.~\ref{fig3}b by blue and red lines, respectively. These density functions, especially the red curve for the LCS significantly differ from the light brown universal density function, $\rho^\mathrm{RMT}(\tau)$. This significant difference is also demonstrated on the level of the ScS $I(V)$ curves in Fig.~\ref{fig3}c. The brown curve shows the reference average $I(V)$ curve corresponding to $\rho^\mathrm{RMT}(\tau)$ such that the light brown region marks the standard deviation of the $I(V)$ curve for different random number transmission sets. Similarly, the light blue and light red regions for the blue and red model $I(V)$ curves demonstrate the standard deviation of these $I(V)$ curves for different random number sets, though for the red curve this standard deviation is so small, that it is not visible on the curve. It is clear, that the deviation of the HCS and LCS $I(V)$ traces from the reference brown curve calculated with $\rho^\mathrm{RMT}(\tau)$ is significantly larger than the standard deviation of the model $I(V)$ curves.

\begin{figure}[t!]
	\centering
	\includegraphics[width=0.475\textwidth]{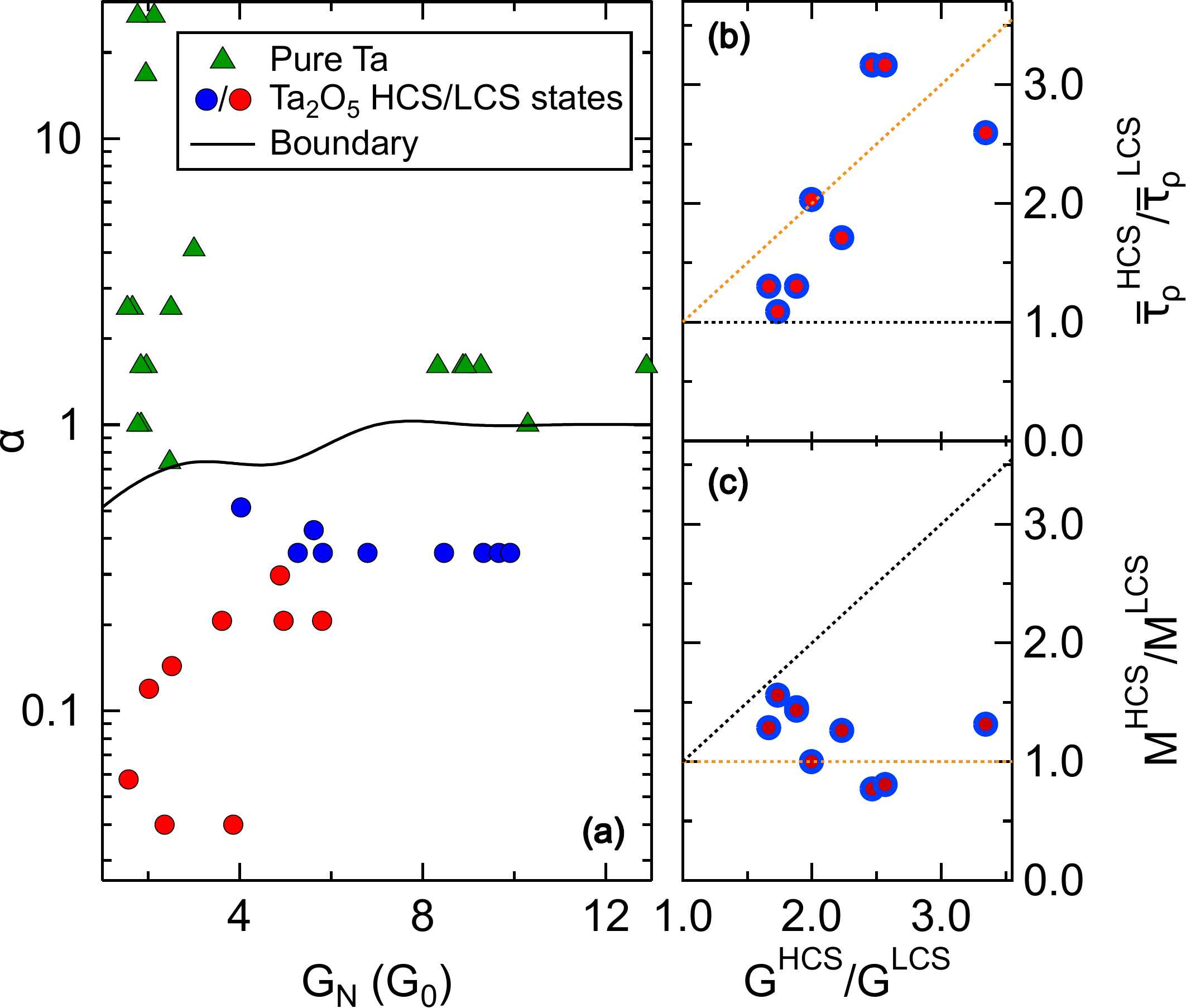}
	\caption{\it \textbf{Analysis of the barrier strength for a statistical ensemble of pure Ta nanowires and RS filaments.} (a) The $\alpha$ values extracted from the ScS $I(V)$ curve fitting are presented for multiple pure Ta atomic wires (green triangles) as well as the HCS (blue dots) and LCS (red dots) states of various RS filaments. The black line demonstrates the border, above/below which the deviation from the $\rho^\mathrm{RMT}(\tau)$ universal transmission density function is insignificant/significant. Panels (b) and (c) respectively demonstrate the average transmission ratio and the channel number ratio as a function of the conductance ratio for all the RS transitions in panel (a). The orange/black dotted line illustrate the cases, where solely the transmission/channel number changes during reset.}
	\label{fig4}
\end{figure}

Figure~\ref{fig4}a displays the results of the statistical analysis performed for a larger amount of oxide-based RS filaments (blue circles for the HCS and red circles for the LCS) as well as for several pure Ta atomic wires (green triangles) covering the conductance range of $\approx 1-13\,$G$_0$. The $\alpha$ parameter of the model density functions is extracted by fitting the ScS $I(V)$ curves likewise in Fig.~\ref{fig3}. The black curve represents a quantitative boundary as a function of the conductance. Below this boundary the standard deviation of the best fitting model $I(V)$ curve does not overlap with the standard deviation of the model $I(V)$ curve corresponding to $\rho^\mathrm{RMT}(\tau)$. That is, below this boundary line the density function best describing the measured $ScS$ $I(V)$ curve significantly differs from $\rho^\mathrm{RMT}(\tau)$. Figure~\ref{fig4} evidences that all the measurements on pure Ta atomic wires / oxide-based RS filaments show non-significant / significant difference compared to the $\rho^\mathrm{RMT}(\tau)$ density function. Furthermore, it is also clear that in the LCS of the RS filaments (red circles) the deviation from $\rho^\mathrm{RMT}(\tau)$ is even more pronounced than in the HCS (blue circles). 

\textit{\textbf{Discussion.}} Finally, we discuss the results of our superconducting subgap analysis. First, we have confirmed our preconception, that the quantum transport in pure Ta atomic-sized nanowires is well described by $\rho^\mathrm{RMT}(\tau)$. Moreover, this universal transmission density is valid in the entire investigated conductance range (see Fig.~\ref{fig4}). As the Ta wire is elongated, the number of the open channels decreases but the characteristic distribution of the transmission eigenvalues and, thereby, the $\overline{\tau_\rho}\approx 0.39$ average transmission is left similar. The $\rho^\mathrm{RMT}(\tau)$ density function represents a broad variation of the transmission eigenvalues, with some highly open channels. We attribute the latter to the transmission through $s$-type orbitals, whereas the $d$-type orbitals rather contribute to the broad variation of $\tau_i$. Despite the few open channels, $\rho^\mathrm{RMT}(\tau)$ does not yield any sign of conductance quantization features, as demonstrated by the simulated conductance histogram in Fig.~S4 of the SI. This is in sharp contrast to noble metals and alkali metals, where a significantly more transmissive transmission density function~\cite{Vardimon2016} is realized due to the dominance of the highly delocalized $s$-electrons and, thus, conductance quantization is observed.\cite{Scheer-Cuevas,Scheer1998}

In filamentary RS devices the formation of atomic-sized metallic filaments is anticipated, where quantum transport properties might resemble those of pure metallic nanowires. Our analysis, however, highlights significant differences. The superconducting excess current\cite{Senkpiel2022} and the leading subgap contribution to the ScS $I(V)$ curve are quadratically suppressed for $\tau \ll 1$, i.e. these superconducting features are extremely sensitive to the presence of transparent channels, even if they represent a minority in the entire distribution. This enables the precise identification of a weak barrier described by $\alpha$, from which the average transmission values and the corresponding number of open channels can be determined. For the representative switching curve in Fig.~\ref{fig3}a,c this analysis yields $\overline{\tau_\rho}=0.25$ and $M=16$ in the HCS and $\overline{\tau_\rho}=0.12$ and $M=17$ in the LCS. These results evidence a markedly different scenario compared to pure Ta nanowires: in the RS filament the reset transition leaves the number of open channels nearly unchanged. Instead, the resistance change occurs due to a significant decrease in the average transmission. Such a process is consistent with the atomistic picture of the RS mechanism where the electric field driven redistribution of oxygen ions at the filament bottleneck (see the schematic illustration in Fig.~\ref{fig2}e) results in a stronger / weaker transport barrier in the LCS / HCS whereas the overall diameter of the filament stays similar. The unchanged filament diameter and modulated oxygen vacancy concentration upon the reset transition is also in qualitative agreement with continuum model simulations performed in the limit of nanometer-scale RS filaments in TaO$_{x}$-based memristors.\cite{Marchewka2016} Our analysis yields an estimate on the filament size assuming that the $5$ channels between two neighboring Ta atoms is a reasonable assumption for the oxide-based RS filaments as well. According to this assumption, the HCS and LCS in Fig.~\ref{fig3}a,c correspond to an RS filament with $\approx 3$ Ta atoms in the narrowest cross-section, with smaller/larger oxygen content in the HCS/LCS bottleneck. 

To perform similar analysis on various RS junctions, we plot the $\overline{\tau}_\rho^\mathrm{HCS}/\overline{\tau}_\rho^\mathrm{LCS}$ (Fig.~\ref{fig4}b) and the $M^\mathrm{HCS}/M^\mathrm{LCS}$ (Fig.~\ref{fig4}c) ratios as a function of the $G^\mathrm{HCS}/G^\mathrm{LCS}$ conductance ratio for the RS data of Fig.~\ref{fig4}a. The orange/black dotted lines illustrate the cases, where solely the transmission/solely the channel number changes along the RS. It is clear that the former process is dominant in the data, i.e. the reset transition leaves the junction size similar, and rather the transmission decreases significantly due to the formation of an oxygen-based barrier at the bottleneck. Relying on the extracted $M$ values the conductance regime displayed in Fig.~\ref{fig4}a can be attributed to RS filaments composed of 3\,--\,8 Ta atoms in their narrowest cross-section.


\section{Conclusions}
In conclusion, we utilized superconducting subgap spectroscopy to reveal the microscopic structure of atomic-sized oxide-based RS filaments in their high and low conductance states. We experimentally determined the transmission density function of the open conduction channels supported by the filamentary region. Our results were analyzed in comparison to the reference system of pure Ta atomic wires. In the latter case the conductance histogram lacks any conductance quantization features due to the presence of partially transmitting $d$ channels. Instead, the transmission density function exhibits a broad distribution which is well described by the universal $\rho^\mathrm{RMT}(\tau)$ density function. The presence of some highly transparent channels is characteristic to this universal transmission distribution, which is sensitively detected by superconducting subgap spectroscopy. In sharp contrast, the highly transparent conduction channels in RS filaments are blocked upon the reset transition. This implies that during reset an extended potential barrier arises at the narrowest cross-section of the filament due to the voltage-induced accumulation of oxygen ions. Meanwhile the filament diameter is preserved at an estimated configuration involving a few Ta atoms. This finding demonstrates that the reset process is fundamentally different from the atom-by-atom thinning of metallic nanowires, resulting in markedly different quantum transport properties. Furthermore, our analysis demonstrates the merits of quantum transport measurements in the microscopic understanding of atomistic processes in filamentary resistive switching devices.

\section*{Author Contributions}
T. N. T. performed the RS measurements and all the data analysis. T. N. T. and P. M. contributed to the measurements on pure Ta atomic wires. M. C. prepared the Ta$_2$O$_5$ layer. Z. B. contributed to MCBJ sample preparation and supervised the low-temperature measurements. A. H. conceived the idea of the project and supervised the project. The manuscript was prepared by A. H., T. N. T. and M. C. All authors contributed to the discussion of the results and the manuscript.  

\section*{Conflicts of interest}
There are no conflicts to declare.

\section*{Acknowledgements}
This research was supported by the Ministry of Culture and Innovation and the National Research, Development and Innovation Office within the Quantum Information National Laboratory of Hungary (Grant No. 2022-2.1.1-NL-2022-00004), and the NKFI K119797 and K143169 grants. T. N. T. acknowledges the support of the KDP-2020 963575 Cooperative Doctoral Programme of the Ministry for Innovation and Technology. Z.B. acknowledges the support of the Bolyai János Research Scholarship of the Hungarian Academy of Sciences and the ÚNKP-22-5 New National Excellence Program of the Ministry for Innovation and Technology from the source of the National Research, Development and Innovation Fund.  The cryogenic measurements were supported by the helium liquefaction plant of the BUTE (VEKOP 2.3.3-15-2017-00015). The authors are grateful to G. Zar\'{a}nd for useful discussions and to G. Rubio-Bollinger for the fitting program based on the theory of multiple Andreev reflections.

\bibliographystyle{achemso}
\bibliography{References}

\providecommand{\latin}[1]{#1}
\makeatletter
\providecommand{\doi}
  {\begingroup\let\do\@makeother\dospecials
  \catcode`\{=1 \catcode`\}=2 \doi@aux}
\providecommand{\doi@aux}[1]{\endgroup\texttt{#1}}
\makeatother
\providecommand*\mcitethebibliography{\thebibliography}
\csname @ifundefined\endcsname{endmcitethebibliography}
  {\let\endmcitethebibliography\endthebibliography}{}
\begin{mcitethebibliography}{38}
\providecommand*\natexlab[1]{#1}
\providecommand*\mciteSetBstSublistMode[1]{}
\providecommand*\mciteSetBstMaxWidthForm[2]{}
\providecommand*\mciteBstWouldAddEndPuncttrue
  {\def\EndOfBibitem{\unskip.}}
\providecommand*\mciteBstWouldAddEndPunctfalse
  {\let\EndOfBibitem\relax}
\providecommand*\mciteSetBstMidEndSepPunct[3]{}
\providecommand*\mciteSetBstSublistLabelBeginEnd[3]{}
\providecommand*\EndOfBibitem{}
\mciteSetBstSublistMode{f}
\mciteSetBstMaxWidthForm{subitem}{(\alph{mcitesubitemcount})}
\mciteSetBstSublistLabelBeginEnd
  {\mcitemaxwidthsubitemform\space}
  {\relax}
  {\relax}

\bibitem[Yang \latin{et~al.}(2013)Yang, Strukov, and Stewart]{Yang2013}
Yang,~J.~J.; Strukov,~D.~B.; Stewart,~D.~R. Memristive devices for computing.
  \emph{Nat. Nanotechnol.} \textbf{2013}, \emph{8}, 13--24\relax
\mciteBstWouldAddEndPuncttrue
\mciteSetBstMidEndSepPunct{\mcitedefaultmidpunct}
{\mcitedefaultendpunct}{\mcitedefaultseppunct}\relax
\EndOfBibitem
\bibitem[Zidan \latin{et~al.}(2018)Zidan, Strachan, and Lu]{Zidan2018}
Zidan,~M.~A.; Strachan,~J.~P.; Lu,~W.~D. The future of electronics based on
  memristive systems. \emph{Nat. Electron.} \textbf{2018}, \emph{1},
  22--29\relax
\mciteBstWouldAddEndPuncttrue
\mciteSetBstMidEndSepPunct{\mcitedefaultmidpunct}
{\mcitedefaultendpunct}{\mcitedefaultseppunct}\relax
\EndOfBibitem
\bibitem[Ambrogio \latin{et~al.}(2018)Ambrogio, Narayanan, Tsai, Shelby,
  Boybat, di~Nolfo, Sidler, Giordano, Bodini, Farinha, Killeen, Cheng, Jaoudi,
  and Burr]{Ambrogio2018}
Ambrogio,~S.; Narayanan,~P.; Tsai,~H.; Shelby,~R.~M.; Boybat,~I.; di~Nolfo,~C.;
  Sidler,~S.; Giordano,~M.; Bodini,~M.; Farinha,~N. C.~P.; Killeen,~B.;
  Cheng,~C.; Jaoudi,~Y.; Burr,~G.~W. Equivalent-accuracy accelerated
  neural-network training using analogue memory. \emph{Nature} \textbf{2018},
  \emph{558}, 60--67\relax
\mciteBstWouldAddEndPuncttrue
\mciteSetBstMidEndSepPunct{\mcitedefaultmidpunct}
{\mcitedefaultendpunct}{\mcitedefaultseppunct}\relax
\EndOfBibitem
\bibitem[Xia and Yang(2019)Xia, and Yang]{Xia2019}
Xia,~Q.; Yang,~J.~J. Memristive crossbar arrays for brain-inspired computing.
  \emph{Nat. Mater.} \textbf{2019}, \emph{18}, 309--323\relax
\mciteBstWouldAddEndPuncttrue
\mciteSetBstMidEndSepPunct{\mcitedefaultmidpunct}
{\mcitedefaultendpunct}{\mcitedefaultseppunct}\relax
\EndOfBibitem
\bibitem[T\"{o}r\"{o}k \latin{et~al.}(2020)T\"{o}r\"{o}k, Csontos, Makk, and
  Halbritter]{Torok2020}
T\"{o}r\"{o}k,~T.~N.; Csontos,~M.; Makk,~P.; Halbritter,~A. Breaking the
  Quantum {PIN} Code of Atomic Synapses. \emph{Nano Lett.} \textbf{2020},
  \emph{20}, 1192--1200\relax
\mciteBstWouldAddEndPuncttrue
\mciteSetBstMidEndSepPunct{\mcitedefaultmidpunct}
{\mcitedefaultendpunct}{\mcitedefaultseppunct}\relax
\EndOfBibitem
\bibitem[Milano \latin{et~al.}(2022)Milano, Aono, Boarino, Celano, Hasegawa,
  Kozicki, Majumdar, Menghini, Miranda, Ricciardi, Tappertzhofen, Terabe, and
  Valov]{Milano2022}
Milano,~G.; Aono,~M.; Boarino,~L.; Celano,~U.; Hasegawa,~T.; Kozicki,~M.;
  Majumdar,~S.; Menghini,~M.; Miranda,~E.; Ricciardi,~C.; Tappertzhofen,~S.;
  Terabe,~K.; Valov,~I. Quantum Conductance in Memristive Devices:
  Fundamentals, Developments, and Applications (Adv. Mater. 32/2022).
  \emph{Adv. Mater.} \textbf{2022}, \emph{34}, 2270235\relax
\mciteBstWouldAddEndPuncttrue
\mciteSetBstMidEndSepPunct{\mcitedefaultmidpunct}
{\mcitedefaultendpunct}{\mcitedefaultseppunct}\relax
\EndOfBibitem
\bibitem[Aono and Hasegawa(2010)Aono, and Hasegawa]{Aono2010}
Aono,~M.; Hasegawa,~T. The Atomic Switch. \emph{IEEE Proc.} \textbf{2010},
  \emph{98}, 2228--2236\relax
\mciteBstWouldAddEndPuncttrue
\mciteSetBstMidEndSepPunct{\mcitedefaultmidpunct}
{\mcitedefaultendpunct}{\mcitedefaultseppunct}\relax
\EndOfBibitem
\bibitem[Cheng \latin{et~al.}(2019)Cheng, Emboras, Salamin, Ducry, Ma,
  Fedoryshyn, Andermatt, Luisier, and Leuthold]{Cheng2019}
Cheng,~B.; Emboras,~A.; Salamin,~Y.; Ducry,~F.; Ma,~P.; Fedoryshyn,~Y.;
  Andermatt,~S.; Luisier,~M.; Leuthold,~J. Ultra compact electrochemical
  metallization cells offering reproducible atomic scale memristive switching.
  \emph{Communications Physics} \textbf{2019}, \emph{2}, 28\relax
\mciteBstWouldAddEndPuncttrue
\mciteSetBstMidEndSepPunct{\mcitedefaultmidpunct}
{\mcitedefaultendpunct}{\mcitedefaultseppunct}\relax
\EndOfBibitem
\bibitem[van Ruitenbeek \latin{et~al.}(2016)van Ruitenbeek, Morales-Masis, and
  Miranda]{res-switch-book}
van Ruitenbeek,~J.~M.; Morales-Masis,~M.; Miranda,~E. In \emph{{Resistive
  Switching: From Fundamentals of Nanoionic Redox Processes to Memristive
  Device Applications Ch. 7. Quantum Point Contact Conduction}}; Ielmini,~D.,
  Waser,~R., Eds.; Wiley-VCH, Weinheim, 2016; pp 197--224\relax
\mciteBstWouldAddEndPuncttrue
\mciteSetBstMidEndSepPunct{\mcitedefaultmidpunct}
{\mcitedefaultendpunct}{\mcitedefaultseppunct}\relax
\EndOfBibitem
\bibitem[Kittel(2005)]{Kittel2005}
Kittel,~C. \emph{Introduction to Solid State Physics, 8$^{th}$ edition}; John
  Wiley \& Sons, Inc., New York, 2005\relax
\mciteBstWouldAddEndPuncttrue
\mciteSetBstMidEndSepPunct{\mcitedefaultmidpunct}
{\mcitedefaultendpunct}{\mcitedefaultseppunct}\relax
\EndOfBibitem
\bibitem[van Wees \latin{et~al.}(1988)van Wees, van Houten, Beenakker,
  Williamson, Kouwenhoven, van~der Marel, and Foxon]{vanWees1988}
van Wees,~B.~J.; van Houten,~H.; Beenakker,~C. W.~J.; Williamson,~J.~G.;
  Kouwenhoven,~L.~P.; van~der Marel,~D.; Foxon,~C.~T. Quantized conductance of
  point contacts in a two-dimensional electron gas. \emph{Phys. Rev. Lett.}
  \textbf{1988}, \emph{60}, 848--850\relax
\mciteBstWouldAddEndPuncttrue
\mciteSetBstMidEndSepPunct{\mcitedefaultmidpunct}
{\mcitedefaultendpunct}{\mcitedefaultseppunct}\relax
\EndOfBibitem
\bibitem[Agra\"it \latin{et~al.}(2003)Agra\"it, Yeyati, and van
  Ruitenbeek]{Ruitenbeek-PhysRep-quantum}
Agra\"it,~N.; Yeyati,~A.~L.; van Ruitenbeek,~J.~M. Quantum Properties of
  Atomic-Sized Conductors. \emph{Phys. Rep.} \textbf{2003}, \emph{377},
  81--279\relax
\mciteBstWouldAddEndPuncttrue
\mciteSetBstMidEndSepPunct{\mcitedefaultmidpunct}
{\mcitedefaultendpunct}{\mcitedefaultseppunct}\relax
\EndOfBibitem
\bibitem[Cuevas and Scheer(2010)Cuevas, and Scheer]{Scheer-Cuevas}
Cuevas,~J.~C.; Scheer,~E. \emph{{Molecular electronics: an introduction to
  theory and experiment}}; World Scientific Series in Nanoscience and
  Nanotechnology; World Scientific: Singapore, 2010\relax
\mciteBstWouldAddEndPuncttrue
\mciteSetBstMidEndSepPunct{\mcitedefaultmidpunct}
{\mcitedefaultendpunct}{\mcitedefaultseppunct}\relax
\EndOfBibitem
\bibitem[Scheer \latin{et~al.}(1998)Scheer, Agra\"it, Cuevas, Yeyati, Ludoph,
  Mart\'in-Rodero, Bollinger, van Ruitenbeek, and Urbina]{Scheer1998}
Scheer,~E.; Agra\"it,~N.; Cuevas,~J.~C.; Yeyati,~A.~L.; Ludoph,~B.;
  Mart\'in-Rodero,~A.; Bollinger,~G.~R.; van Ruitenbeek,~J.~M.; Urbina,~C. The
  signature of chemical valence in the electrical conduction through a
  single-atom contact. \emph{Nature} \textbf{1998}, \emph{394}, 154--157\relax
\mciteBstWouldAddEndPuncttrue
\mciteSetBstMidEndSepPunct{\mcitedefaultmidpunct}
{\mcitedefaultendpunct}{\mcitedefaultseppunct}\relax
\EndOfBibitem
\bibitem[Park \latin{et~al.}(2013)Park, Kim, Park, Li, Heo, Lee, Chang, Kwon,
  Kim, Chung, Dittmann, Waser, and Kim]{Park2013}
Park,~G.-S.; Kim,~Y.~B.; Park,~S.~Y.; Li,~X.~S.; Heo,~S.; Lee,~M.-J.;
  Chang,~M.; Kwon,~J.~H.; Kim,~M.; Chung,~U.-I.; Dittmann,~R.; Waser,~R.;
  Kim,~K. In situ observation of filamentary conducting channels in an
  asymmetric Ta$_{2}$O$_{5-x}$/TaO$_{2-x}$ bilayer structure. \emph{Nat.
  Commun.} \textbf{2013}, \emph{4}, 2382\relax
\mciteBstWouldAddEndPuncttrue
\mciteSetBstMidEndSepPunct{\mcitedefaultmidpunct}
{\mcitedefaultendpunct}{\mcitedefaultseppunct}\relax
\EndOfBibitem
\bibitem[Marchewka \latin{et~al.}(2016)Marchewka, Roesgen, Skaja, Du, Jia,
  Mayer, Rana, Waser, and Menzel]{Marchewka2016}
Marchewka,~A.; Roesgen,~B.; Skaja,~K.; Du,~H.; Jia,~C.-L.; Mayer,~J.; Rana,~V.;
  Waser,~R.; Menzel,~S. Nanoionic Resistive Switching Memories: On the Physical
  Nature of the Dynamic Reset Process. \emph{Adv. El. Mater.} \textbf{2016},
  \emph{2}, 1500233\relax
\mciteBstWouldAddEndPuncttrue
\mciteSetBstMidEndSepPunct{\mcitedefaultmidpunct}
{\mcitedefaultendpunct}{\mcitedefaultseppunct}\relax
\EndOfBibitem
\bibitem[Yang \latin{et~al.}(2010)Yang, Zhang, Strachan, Miao, Pickett, Kelley,
  Medeiros-Ribeiro, and Williams]{Yang2010}
Yang,~J.~J.; Zhang,~M.-X.; Strachan,~J.~P.; Miao,~F.; Pickett,~M.~D.;
  Kelley,~R.~D.; Medeiros-Ribeiro,~G.; Williams,~R.~S. High switching endurance
  in TaOx memristive devices. \emph{Appl. Phys. Lett.} \textbf{2010},
  \emph{97}, 232102\relax
\mciteBstWouldAddEndPuncttrue
\mciteSetBstMidEndSepPunct{\mcitedefaultmidpunct}
{\mcitedefaultendpunct}{\mcitedefaultseppunct}\relax
\EndOfBibitem
\bibitem[Torrezan \latin{et~al.}(2011)Torrezan, Strachan, Medeiros-Ribeiro, and
  Williams]{Torrezan2011}
Torrezan,~A.~C.; Strachan,~J.~W.; Medeiros-Ribeiro,~G.; Williams,~R.~S.
  Sub-nanosecond switching of a tantalum oxide memristor. \emph{Nanotechnology}
  \textbf{2011}, \emph{22}, 485203\relax
\mciteBstWouldAddEndPuncttrue
\mciteSetBstMidEndSepPunct{\mcitedefaultmidpunct}
{\mcitedefaultendpunct}{\mcitedefaultseppunct}\relax
\EndOfBibitem
\bibitem[Strachan \latin{et~al.}(2011)Strachan, Torrezan, Medeiros-Ribeiro, and
  Williams]{Strachan2011}
Strachan,~J.~P.; Torrezan,~A.~C.; Medeiros-Ribeiro,~G.; Williams,~R.~S.
  Measuring the switching dynamics and energy efficiency of tantalum oxide
  memristors. \emph{Nanotechnology} \textbf{2011}, \emph{22}, 505402\relax
\mciteBstWouldAddEndPuncttrue
\mciteSetBstMidEndSepPunct{\mcitedefaultmidpunct}
{\mcitedefaultendpunct}{\mcitedefaultseppunct}\relax
\EndOfBibitem
\bibitem[Lee \latin{et~al.}(2011)Lee, Lee, Lee, Lee, Chang, Hur, Kim, Kim, Seo,
  Seo, Chung, Yoo, and Kim]{Lee2011}
Lee,~M.-J.; Lee,~C.~B.; Lee,~D.; Lee,~S.; Chang,~M.; Hur,~J.; Kim,~Y.-B.;
  Kim,~C.-J.; Seo,~D.; Seo,~S.; Chung,~U.-I.; Yoo,~I.-K.; Kim,~K. A Fast,
  High-Endurance and Scalable Non-Volatile Memory Device Made from Asymmetric
  Ta2O5-X/TaO2-X Bilayer Structures. \emph{Nat. Mater.} \textbf{2011},
  \emph{10}, 625--630\relax
\mciteBstWouldAddEndPuncttrue
\mciteSetBstMidEndSepPunct{\mcitedefaultmidpunct}
{\mcitedefaultendpunct}{\mcitedefaultseppunct}\relax
\EndOfBibitem
\bibitem[Graves \latin{et~al.}(2017)Graves, Dávila, Merced-Grafals, Lam,
  Strachan, and Williams]{Graves2017}
Graves,~C.~E.; Dávila,~N.; Merced-Grafals,~E.~J.; Lam,~S.-T.; Strachan,~J.~P.;
  Williams,~R.~S. Temperature and field-dependent transport measurements in
  continuously tunable tantalum oxide memristors expose the dominant state
  variable. \emph{Appl. Phys. Lett.} \textbf{2017}, \emph{110}, 123501\relax
\mciteBstWouldAddEndPuncttrue
\mciteSetBstMidEndSepPunct{\mcitedefaultmidpunct}
{\mcitedefaultendpunct}{\mcitedefaultseppunct}\relax
\EndOfBibitem
\bibitem[Rao \latin{et~al.}(2022)Rao, Song, Kiani, Asapu, Zhuo, Midya,
  Upadhyay, Wu, Barnell, Lin, Li, .i~Wang, Xia, and Yang]{Rao2022}
Rao,~M.; Song,~W.; Kiani,~F.; Asapu,~S.; Zhuo,~Y.; Midya,~R.; Upadhyay,~N.;
  Wu,~Q.; Barnell,~M.; Lin,~P.; Li,~C.; .i~Wang,; Xia,~Q.; Yang,~J.~J. Timing
  Selector: Using Transient Switching Dynamics to Solve the Sneak Path Issue of
  Crossbar Arrays. \emph{Small Science} \textbf{2022}, \emph{2}, 2100072\relax
\mciteBstWouldAddEndPuncttrue
\mciteSetBstMidEndSepPunct{\mcitedefaultmidpunct}
{\mcitedefaultendpunct}{\mcitedefaultseppunct}\relax
\EndOfBibitem
\bibitem[Csontos \latin{et~al.}(2022)Csontos, Horst, Olalla, Koch, Shorubalko,
  Halbritter, and Leuthold]{Csontos2022}
Csontos,~M.; Horst,~Y.; Olalla,~N.~J.; Koch,~U.; Shorubalko,~I.;
  Halbritter,~A.; Leuthold,~J. Picosecond Time-Scale Resistive Switching
  Monitored in Real-Time. \emph{arXiv:2209.06732} \textbf{2022}, \relax
\mciteBstWouldAddEndPunctfalse
\mciteSetBstMidEndSepPunct{\mcitedefaultmidpunct}
{}{\mcitedefaultseppunct}\relax
\EndOfBibitem
\bibitem[Scheer \latin{et~al.}(1997)Scheer, Joyez, Esteve, Urbina, and
  Devoret]{Scheer1997}
Scheer,~E.; Joyez,~P.; Esteve,~D.; Urbina,~C.; Devoret,~M.~H. Conduction
  Channel Transmissions of Atomic-Size Aluminum Contacts. \emph{Phys. Rev.
  Lett.} \textbf{1997}, \emph{78}, 3535--3538\relax
\mciteBstWouldAddEndPuncttrue
\mciteSetBstMidEndSepPunct{\mcitedefaultmidpunct}
{\mcitedefaultendpunct}{\mcitedefaultseppunct}\relax
\EndOfBibitem
\bibitem[Ludoph \latin{et~al.}(2000)Ludoph, van~der Post, Bratus', Bezuglyi,
  Shumeiko, Wendin, and van Ruitenbeek]{Ludoph2000}
Ludoph,~B.; van~der Post,~N.; Bratus',~E.~N.; Bezuglyi,~E.~V.; Shumeiko,~V.~S.;
  Wendin,~G.; van Ruitenbeek,~J.~M. Multiple {Andreev} reflection in
  single-atom niobium junctions. \emph{Phys. Rev. B: Condens. Matter Mater.
  Phys.} \textbf{2000}, \emph{61}, 8561\relax
\mciteBstWouldAddEndPuncttrue
\mciteSetBstMidEndSepPunct{\mcitedefaultmidpunct}
{\mcitedefaultendpunct}{\mcitedefaultseppunct}\relax
\EndOfBibitem
\bibitem[Makk \latin{et~al.}(2008)Makk, Csonka, and Halbritter]{Makk2008}
Makk,~P.; Csonka,~S.; Halbritter,~A. Effect of hydrogen molecules on the
  electronic transport through atomic-sized metallic junctions in the
  superconducting state. \emph{Phys. Rev. B: Condens. Matter Mater. Phys.}
  \textbf{2008}, \emph{78}, 045414\relax
\mciteBstWouldAddEndPuncttrue
\mciteSetBstMidEndSepPunct{\mcitedefaultmidpunct}
{\mcitedefaultendpunct}{\mcitedefaultseppunct}\relax
\EndOfBibitem
\bibitem[Landauer(1970)]{Landauer1970}
Landauer,~R. Electrical resistance of disordered one-dimensional lattices.
  \emph{Philos. Mag.} \textbf{1970}, \emph{21}, 863--867\relax
\mciteBstWouldAddEndPuncttrue
\mciteSetBstMidEndSepPunct{\mcitedefaultmidpunct}
{\mcitedefaultendpunct}{\mcitedefaultseppunct}\relax
\EndOfBibitem
\bibitem[Beenakker(1997)]{Beenakker1997}
Beenakker,~C. W.~J. Random-matrix theory of quantum transport. \emph{Rev. Mod.
  Phys.} \textbf{1997}, \emph{69}, 731--808\relax
\mciteBstWouldAddEndPuncttrue
\mciteSetBstMidEndSepPunct{\mcitedefaultmidpunct}
{\mcitedefaultendpunct}{\mcitedefaultseppunct}\relax
\EndOfBibitem
\bibitem[Li \latin{et~al.}(2018)Li, Hu, Li, Jiang, Ge, Montgomery, Zhang, Song,
  D\'avila, Graves, Li, Strachan, Lin, Wang, Barnell, Wu, Williams, Yang, and
  Xia]{Li2018}
Li,~C. \latin{et~al.}  Analogue signal and image processing with large
  memristor crossbars. \emph{Nat. Electron.} \textbf{2018}, \emph{1},
  52--59\relax
\mciteBstWouldAddEndPuncttrue
\mciteSetBstMidEndSepPunct{\mcitedefaultmidpunct}
{\mcitedefaultendpunct}{\mcitedefaultseppunct}\relax
\EndOfBibitem
\bibitem[Henny \latin{et~al.}(1999)Henny, Oberholzer, Strunk, and
  Sch\"onenberger]{Henny1999}
Henny,~M.; Oberholzer,~S.; Strunk,~C.; Sch\"onenberger,~C. 1/3-shot-noise
  suppression in diffusive nanowires. \emph{Phys. Rev. B: Condens. Matter
  Mater. Phys.} \textbf{1999}, \emph{59}, 2871--2880\relax
\mciteBstWouldAddEndPuncttrue
\mciteSetBstMidEndSepPunct{\mcitedefaultmidpunct}
{\mcitedefaultendpunct}{\mcitedefaultseppunct}\relax
\EndOfBibitem
\bibitem[Riquelme \latin{et~al.}(2005)Riquelme, de~la Vega, Yeyati, Agra\"it,
  Martin-Rodero, and Rubio-Bollinger]{Rubio-Bollinger-subgap}
Riquelme,~J.~J.; de~la Vega,~L.; Yeyati,~A.~L.; Agra\"it,~N.;
  Martin-Rodero,~A.; Rubio-Bollinger,~G. Distribution of conduction channels in
  nanoscale contacts: Evolution towards the diffusive limit. \emph{Europhysics
  Letters (EPL)} \textbf{2005}, \emph{70}, 663--669\relax
\mciteBstWouldAddEndPuncttrue
\mciteSetBstMidEndSepPunct{\mcitedefaultmidpunct}
{\mcitedefaultendpunct}{\mcitedefaultseppunct}\relax
\EndOfBibitem
\bibitem[Cuevas \latin{et~al.}(1998)Cuevas, Yeyati, and
  Mart\'{\i}n-Rodero]{Cuevas_PRL_1998}
Cuevas,~J.~C.; Yeyati,~A.~L.; Mart\'{\i}n-Rodero,~A. Microscopic Origin of
  Conducting Channels in Metallic Atomic-Size Contacts. \emph{Phys. Rev. Lett.}
  \textbf{1998}, \emph{80}, 1066--1069\relax
\mciteBstWouldAddEndPuncttrue
\mciteSetBstMidEndSepPunct{\mcitedefaultmidpunct}
{\mcitedefaultendpunct}{\mcitedefaultseppunct}\relax
\EndOfBibitem
\bibitem[Nazarov(1994)]{Nazarov1994}
Nazarov,~Y.~V. Limits of universality in disordered conductors. \emph{Phys.
  Rev. Lett.} \textbf{1994}, \emph{73}, 134--137\relax
\mciteBstWouldAddEndPuncttrue
\mciteSetBstMidEndSepPunct{\mcitedefaultmidpunct}
{\mcitedefaultendpunct}{\mcitedefaultseppunct}\relax
\EndOfBibitem
\bibitem[Cuevas \latin{et~al.}(1996)Cuevas, Mart\'in-Rodero, and
  Levy~Yeyati]{Cuevas1996}
Cuevas,~J.~C.; Mart\'in-Rodero,~A.; Levy~Yeyati,~A. Hamiltonian approach to the
  transport properties of superconducting quantum point contacts. \emph{Phys.
  Rev. B: Condens. Matter Mater. Phys.} \textbf{1996}, \emph{54},
  7366--7379\relax
\mciteBstWouldAddEndPuncttrue
\mciteSetBstMidEndSepPunct{\mcitedefaultmidpunct}
{\mcitedefaultendpunct}{\mcitedefaultseppunct}\relax
\EndOfBibitem
\bibitem[Averin and Bardas(1995)Averin, and Bardas]{Averin1995}
Averin,~D.; Bardas,~A. ac Josephson Effect in a Single Quantum Channel.
  \emph{Phys. Rev. Lett.} \textbf{1995}, \emph{75}, 1831--1834\relax
\mciteBstWouldAddEndPuncttrue
\mciteSetBstMidEndSepPunct{\mcitedefaultmidpunct}
{\mcitedefaultendpunct}{\mcitedefaultseppunct}\relax
\EndOfBibitem
\bibitem[Senkpiel \latin{et~al.}(2022)Senkpiel, Drost, Kl\"ockner, Etzkorn,
  Ankerhold, Cuevas, Pauly, Kern, and Ast]{Senkpiel2022}
Senkpiel,~J.; Drost,~R.; Kl\"ockner,~J.~C.; Etzkorn,~M.; Ankerhold,~J.;
  Cuevas,~J.~C.; Pauly,~F.; Kern,~K.; Ast,~C.~R. Extracting transport channel
  transmissions in scanning tunneling microscopy using superconducting excess
  current. \emph{Phys. Rev. B: Condens. Matter Mater. Phys.} \textbf{2022},
  \emph{105}, 165401\relax
\mciteBstWouldAddEndPuncttrue
\mciteSetBstMidEndSepPunct{\mcitedefaultmidpunct}
{\mcitedefaultendpunct}{\mcitedefaultseppunct}\relax
\EndOfBibitem
\bibitem[Levy and Rudnick(1963)Levy, and Rudnick]{Levy1963}
Levy,~M.; Rudnick,~I. Ultrasonic Determination of the Superconducting Energy
  Gap in Tantalum. \emph{Phys. Rev.} \textbf{1963}, \emph{132},
  1073--1080\relax
\mciteBstWouldAddEndPuncttrue
\mciteSetBstMidEndSepPunct{\mcitedefaultmidpunct}
{\mcitedefaultendpunct}{\mcitedefaultseppunct}\relax
\EndOfBibitem
\bibitem[Vardimon \latin{et~al.}(2016)Vardimon, Matt, Nielaba, Cuevas, and
  Tal]{Vardimon2016}
Vardimon,~R.; Matt,~M.; Nielaba,~P.; Cuevas,~J.~C.; Tal,~O. Orbital origin of
  the electrical conduction in ferromagnetic atomic-size contacts: Insights
  from shot noise measurements and theoretical simulations. \emph{Phys. Rev. B:
  Condens. Matter Mater. Phys.} \textbf{2016}, \emph{93}, 085439\relax
\mciteBstWouldAddEndPuncttrue
\mciteSetBstMidEndSepPunct{\mcitedefaultmidpunct}
{\mcitedefaultendpunct}{\mcitedefaultseppunct}\relax
\EndOfBibitem
\end{mcitethebibliography}


\providecommand{\latin}[1]{#1}
\makeatletter
\providecommand{\doi}
  {\begingroup\let\do\@makeother\dospecials
  \catcode`\{=1 \catcode`\}=2 \doi@aux}
\providecommand{\doi@aux}[1]{\endgroup\texttt{#1}}
\makeatother
\providecommand*\mcitethebibliography{\thebibliography}
\csname @ifundefined\endcsname{endmcitethebibliography}
  {\let\endmcitethebibliography\endthebibliography}{}
\begin{mcitethebibliography}{9}
\providecommand*\natexlab[1]{#1}
\providecommand*\mciteSetBstSublistMode[1]{}
\providecommand*\mciteSetBstMaxWidthForm[2]{}
\providecommand*\mciteBstWouldAddEndPuncttrue
  {\def\EndOfBibitem{\unskip.}}
\providecommand*\mciteBstWouldAddEndPunctfalse
  {\let\EndOfBibitem\relax}
\providecommand*\mciteSetBstMidEndSepPunct[3]{}
\providecommand*\mciteSetBstSublistLabelBeginEnd[3]{}
\providecommand*\EndOfBibitem{}
\mciteSetBstSublistMode{f}
\mciteSetBstMaxWidthForm{subitem}{(\alph{mcitesubitemcount})}
\mciteSetBstSublistLabelBeginEnd
  {\mcitemaxwidthsubitemform\space}
  {\relax}
  {\relax}

\bibitem[Gubicza \latin{et~al.}(2016)Gubicza, Manrique, P\'osa, Lambert,
  Mih\'aly, Csontos, and Halbritter]{Gubicza2016}
Gubicza,~A.; Manrique,~D.~Z.; P\'osa,~L.; Lambert,~C.~J.; Mih\'aly,~G.;
  Csontos,~M.; Halbritter,~A. Asymmetry-induced resistive switching in
  {Ag-Ag$_{2}$S-Ag} memristors enabling a simplified atomic-scale memory
  design. \emph{Sci. Rep.} \textbf{2016}, \emph{6}, 30775\relax
\mciteBstWouldAddEndPuncttrue
\mciteSetBstMidEndSepPunct{\mcitedefaultmidpunct}
{\mcitedefaultendpunct}{\mcitedefaultseppunct}\relax
\EndOfBibitem
\bibitem[Cuevas \latin{et~al.}(1996)Cuevas, Mart\'in-Rodero, and
  Levy~Yeyati]{Cuevas1996}
Cuevas,~J.~C.; Mart\'in-Rodero,~A.; Levy~Yeyati,~A. Hamiltonian approach to the
  transport properties of superconducting quantum point contacts. \emph{Phys.
  Rev. B} \textbf{1996}, \emph{54}, 7366--7379\relax
\mciteBstWouldAddEndPuncttrue
\mciteSetBstMidEndSepPunct{\mcitedefaultmidpunct}
{\mcitedefaultendpunct}{\mcitedefaultseppunct}\relax
\EndOfBibitem
\bibitem[Riquelme \latin{et~al.}(2005)Riquelme, de~la Vega, Yeyati, Agra\"it,
  Martin-Rodero, and Rubio-Bollinger]{Rubio-Bollinger-subgap}
Riquelme,~J.~J.; de~la Vega,~L.; Yeyati,~A.~L.; Agra\"it,~N.;
  Martin-Rodero,~A.; Rubio-Bollinger,~G. Distribution of conduction channels in
  nanoscale contacts: Evolution towards the diffusive limit. \emph{Europhysics
  Letters (EPL)} \textbf{2005}, \emph{70}, 663--669\relax
\mciteBstWouldAddEndPuncttrue
\mciteSetBstMidEndSepPunct{\mcitedefaultmidpunct}
{\mcitedefaultendpunct}{\mcitedefaultseppunct}\relax
\EndOfBibitem
\bibitem[Scheer \latin{et~al.}(1998)Scheer, Agra\"it, Cuevas, Yeyati, Ludoph,
  Martín-Rodero, Rubio~Bollinger, Ruitenbeek, and Urbina]{Scheer1998Nature}
Scheer,~E.; Agra\"it,~N.; Cuevas,~J.~C.; Yeyati,~A.~L.; Ludoph,~B.;
  Martín-Rodero,~A.; Rubio~Bollinger,~G.; Ruitenbeek,~J. M.~v.; Urbina,~C. The
  signature of chemical valence in the electrical conduction through a
  single-atom contact. \emph{Nature} \textbf{1998}, \emph{394}, 154--157\relax
\mciteBstWouldAddEndPuncttrue
\mciteSetBstMidEndSepPunct{\mcitedefaultmidpunct}
{\mcitedefaultendpunct}{\mcitedefaultseppunct}\relax
\EndOfBibitem
\bibitem[Török \latin{et~al.}(2020)Török, Csontos, Makk, and
  Halbritter]{Torok2020}
Török,~T.~N.; Csontos,~M.; Makk,~P.; Halbritter,~A. Breaking the Quantum PIN
  Code of Atomic Synapses. \emph{Nano Letters} \textbf{2020}, \emph{20},
  1192--1200\relax
\mciteBstWouldAddEndPuncttrue
\mciteSetBstMidEndSepPunct{\mcitedefaultmidpunct}
{\mcitedefaultendpunct}{\mcitedefaultseppunct}\relax
\EndOfBibitem
\bibitem[Scheer \latin{et~al.}(2001)Scheer, Belzig, Naveh, Devoret, Esteve, and
  Urbina]{Proximity-Scheer}
Scheer,~E.; Belzig,~W.; Naveh,~Y.; Devoret,~M.~H.; Esteve,~D.; Urbina,~C.
  Proximity Effect and Multiple {Andreev} Reflections in Gold Atomic Contacts.
  \emph{Phys. Rev. Lett.} \textbf{2001}, \emph{86}, 284--287\relax
\mciteBstWouldAddEndPuncttrue
\mciteSetBstMidEndSepPunct{\mcitedefaultmidpunct}
{\mcitedefaultendpunct}{\mcitedefaultseppunct}\relax
\EndOfBibitem
\bibitem[Senkpiel \latin{et~al.}(2022)Senkpiel, Drost, Kl\"ockner, Etzkorn,
  Ankerhold, Cuevas, Pauly, Kern, and Ast]{Senkpiel2022}
Senkpiel,~J.; Drost,~R.; Kl\"ockner,~J.~C.; Etzkorn,~M.; Ankerhold,~J.;
  Cuevas,~J.~C.; Pauly,~F.; Kern,~K.; Ast,~C.~R. Extracting transport channel
  transmissions in scanning tunneling microscopy using superconducting excess
  current. \emph{Phys. Rev. B: Condens. Matter Mater. Phys.} \textbf{2022},
  \emph{105}, 165401\relax
\mciteBstWouldAddEndPuncttrue
\mciteSetBstMidEndSepPunct{\mcitedefaultmidpunct}
{\mcitedefaultendpunct}{\mcitedefaultseppunct}\relax
\EndOfBibitem
\bibitem[Cuevas \latin{et~al.}(1998)Cuevas, Yeyati, and
  Mart\'{\i}n-Rodero]{Cuevas_PRL_1998}
Cuevas,~J.~C.; Yeyati,~A.~L.; Mart\'{\i}n-Rodero,~A. Microscopic Origin of
  Conducting Channels in Metallic Atomic-Size Contacts. \emph{Phys. Rev. Lett.}
  \textbf{1998}, \emph{80}, 1066--1069\relax
\mciteBstWouldAddEndPuncttrue
\mciteSetBstMidEndSepPunct{\mcitedefaultmidpunct}
{\mcitedefaultendpunct}{\mcitedefaultseppunct}\relax
\EndOfBibitem
\bibitem[Cuevas and Scheer(2010)Cuevas, and Scheer]{Scheer-Cuevas}
Cuevas,~J.~C.; Scheer,~E. \emph{{Molecular electronics: an introduction to
  theory and experiment}}; World Scientific Series in Nanoscience and
  Nanotechnology; World Scientific: Singapore, 2010\relax
\mciteBstWouldAddEndPuncttrue
\mciteSetBstMidEndSepPunct{\mcitedefaultmidpunct}
{\mcitedefaultendpunct}{\mcitedefaultseppunct}\relax
\EndOfBibitem
\end{mcitethebibliography}

\end{document}


\title{Quantum transport properties of tantalum-oxide resistive switching filaments\\Supplementary information}

\author{Tímea Nóra Török}
\affiliation{Department of Physics, Institute of Physics, Budapest University of Technology and Economics, M\H{u}egyetem rkp. 3., H-1111 Budapest, Hungary.}
\affiliation{Institute of Technical Physics and Materials Science, Centre for Energy Research, Konkoly-Thege M. \'{u}t 29-33, 1121 Budapest, Hungary.}

\author{Péter Makk}
\affiliation{Department of Physics, Institute of Physics, Budapest University of Technology and Economics, M\H{u}egyetem rkp. 3., H-1111 Budapest, Hungary.}
\affiliation{MTA-BME Correlated van der Waals Structures Momentum Research Group, M\H{u}egyetem rkp. 3., H-1111 Budapest, Hungary.}

\author{Zoltán Balogh}
\affiliation{Department of Physics, Institute of Physics, Budapest University of Technology and Economics, M\H{u}egyetem rkp. 3., H-1111 Budapest, Hungary.}
\affiliation{ELKH-BME Condensed Matter Research Group, M\H{u}egyetem rkp. 3., H-1111 Budapest, Hungary.}

\author{Miklós Csontos}
\affiliation{Institute of Electromagnetic Fields, ETH Zurich, Gloriastrasse 35, 8092 Zurich, Switzerland.}

\author{András Halbritter$^{\ast}$}
\affiliation{Department of Physics, Institute of Physics, Budapest University of Technology and Economics, M\H{u}egyetem rkp. 3., H-1111 Budapest, Hungary.}
\affiliation{ELKH-BME Condensed Matter Research Group, M\H{u}egyetem rkp. 3., H-1111 Budapest, Hungary.}

\maketitle

\section{Resistive switching measurements with Ta STM tips}

\begin{figure}[h!]
		\includegraphics[width=0.7\linewidth]{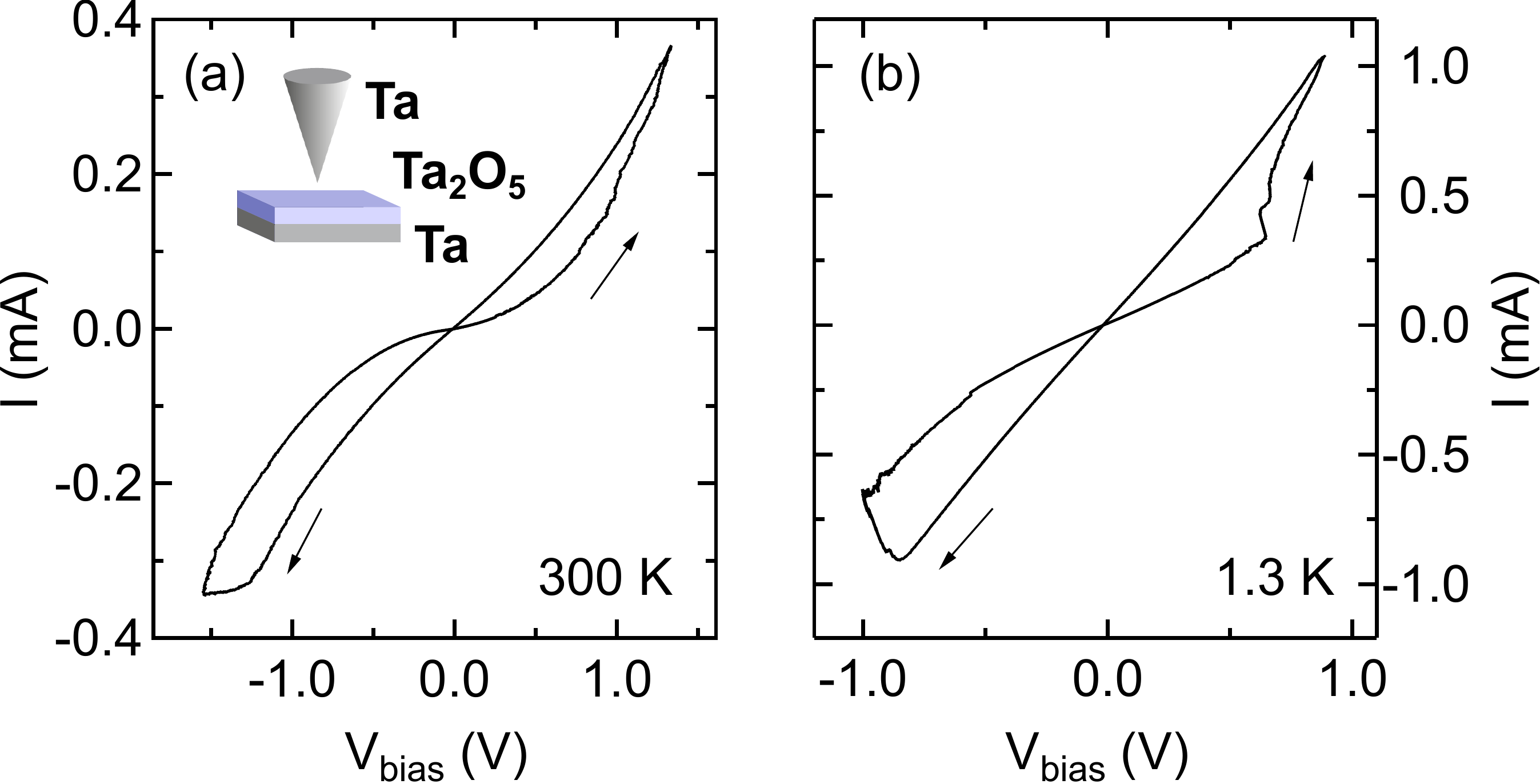}
		\caption{\textbf{Typical resistive switching characteristics of Ta(tip)/Ta$_2$O$_5$/Ta junctions} (a) at $T=300\,$K, (b) at $T=1.3\,$K. The arrows indicate the direction of the hysteresis. The inset in (a) illustrates the STM arrangement.}
		\label{Fig1}
\end{figure}

Resistive switching (RS) $I(V)$ characteristics of the Ta$_2$O$_5$/Ta thin films have been routinely recorded at room temperature utilizing PtIr STM tips as top electrodes. However, the application of superconducting subgap spectroscopy required the application of superconducting electrodes on both side at cryogenic temperature. Figure~\ref{Fig1} demonstrates that similar RS characteristics can be acquired both at room temperature and at $T=1.3\,$K when the PtIr tip is replaced with Ta, despite the material symmetry of the electrodes. The voltage polarity of the set/reset switchings is determined by the geometrical asymmetry of the electrodes which induces a highly asymmetric potential profile during switching events \cite{Gubicza2016}.

\newpage
\section{Multilevel programming in the vicinity of the quantum conductance unit}

\begin{figure}[h!]
		\includegraphics[width=0.6\linewidth]{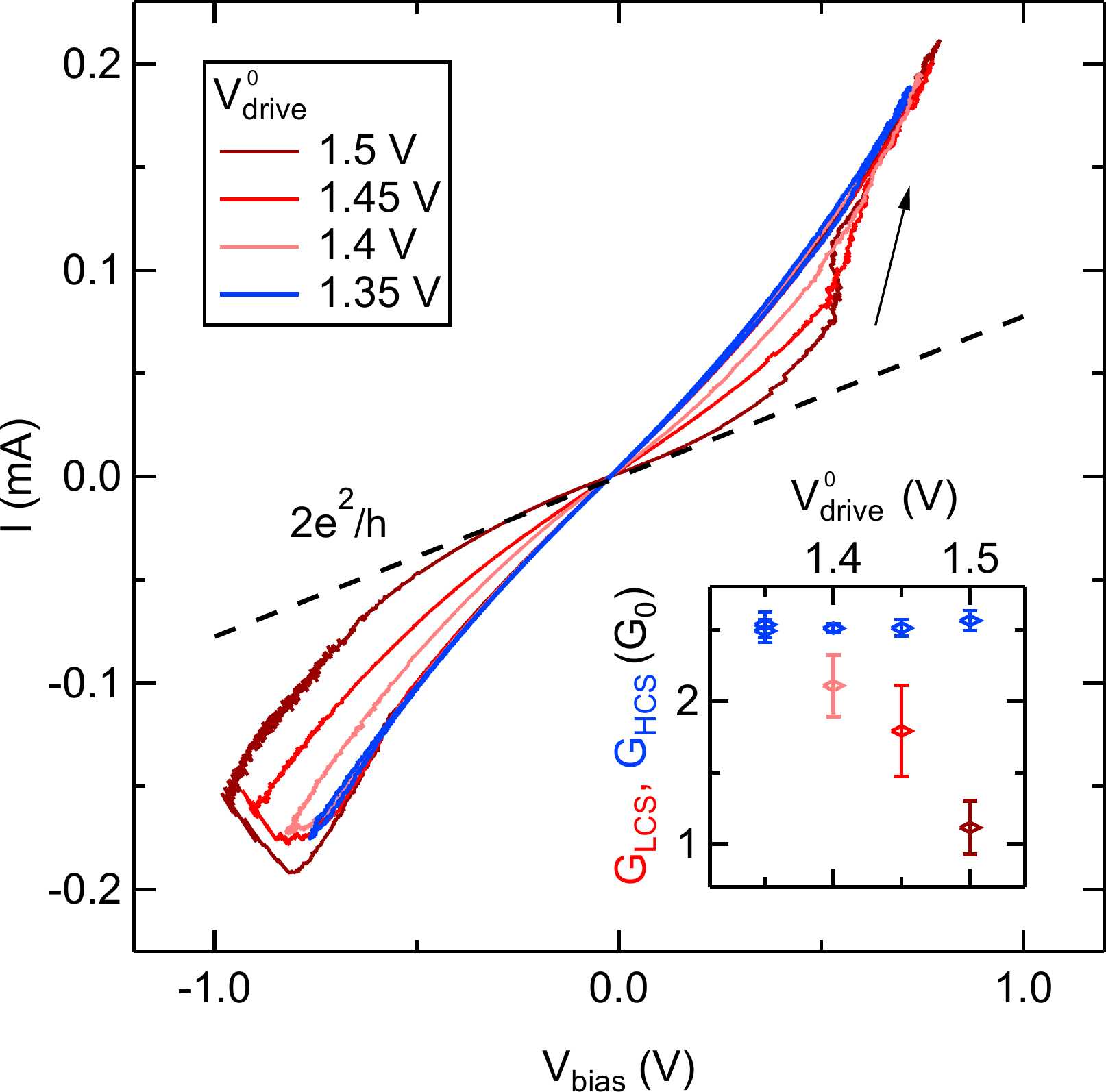}
		\caption{\textbf{Multilevel resistive switching in PtIr/Ta$_2$O$_5$/Ta devices at room temperature.} Typical driving voltage amplitude dependent $I(V)$ curves of PtIr(tip)/Ta$_2$O$_5$/Ta junctions exhibiting various conductance states in the $2.5-1~G_0$ conductance range. The arrow indicates the direction of the hysteresis. The Inset shows the low-bias conductance of low and high conductance states as a function of the driving voltage amplitude (average and standard deviation of 10 consecutive switching cycles are shown for each point). The dashed line is guide to the eye, its slope corresponds to the $G_{0}=2e^{2}/h$ conductance quantum unit.}
		\label{Fig2}
\end{figure}

The analog tunability of room temperature resistive switching in the vicinity of the $G_{0}=2e^{2}/h$ quantum conductance unit is demonstrated in Figure~\ref{Fig2}. The increase of the $V_{\rm drive}^{0}$ amplitude of the triangular driving voltage signal induced a gradual decrease in the LCS conductance across the $2.5-1~G_0$ conductance range (see the inset of Fig.~\ref{Fig2}) while the HCS was only moderately increased. Reproducible quantized steps were not observed during the RS transitions.

\newpage
\section{Subgap spectra of ScS junctions with and without energy smearing}

\begin{figure}[h!]
		\includegraphics[width=0.7\linewidth]{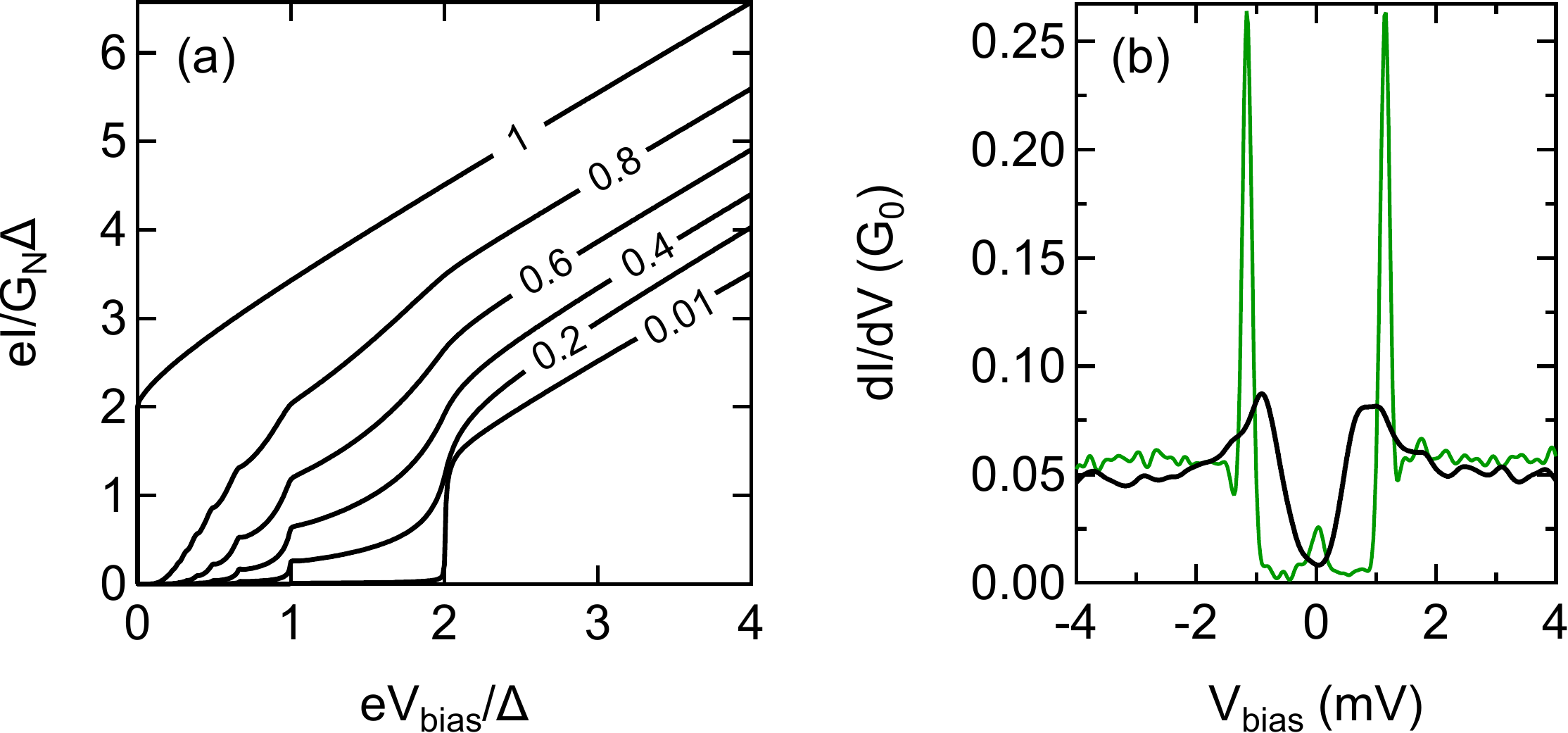}
		\caption{\textbf{Calculated subgap spectra of ScS junctions and the measured differential conductance of ScS tunnel junctions.} (a) Subgap $I(V)$ characteristics of different ScS junctions with various transmission values. (b) Typical differential conductance spectra of pure Ta nanowires (green) and Ta/Ta$_2$O$_5$/Ta RS devices (black) in the tunneling regime. The width of the characteristic peaks at the gap edge is denoted as $\Gamma$ while their distance is $4\Delta$. The parameters of the two curves are the following: $\Delta_{\rm MCBJ}=580~\mu$V and $\Delta_{\rm ox.}=454~\mu$V, $\Gamma_{\rm MCBJ}=185~\mu$V and $\Gamma_{\rm ox.}=600~\mu$V.}
		\label{Fig3}
\end{figure}

Figure~\ref{Fig3}a shows the calculated subgap $I(V)$ characteristics of ScS junctions with a single conductance channel at various transmissions ($\tau=$ 0.01, 0.2, 0.4, 0.6, 0.8, 1, respectively). These theoretically computed fine-structured $I(V)$ characteristics exhibit distinct structures at the fractional values of the superconducting gap,  $eV_{\text{bias}}=2\Delta/n$. Shortly, at $eV_{\text{bias}}<2\Delta/n$ an $n$-th order process with the simultaneous transfer of $n$ electron charges across the constriction becomes available  according to the theory of multiple Andreev reflections \cite{Cuevas1996,Rubio-Bollinger-subgap}. The probability of such an  $n$-th order process, however, scales with $\tau^n$, i.e., these higher order processes respectively yield negligible (significant) subgap current for $\tau\ll 1$ (for a transmission approaching unity). For an ScS junction with an arbitrary $\tau_i$ set of transmission eigenvalues the total $I(V)$ curve is obtained as the sum of the subgap $I(V)$ curves for the various channels. 

The theoretical $I(V)$ curves in Fig.~\ref{Fig3}a well describe the measured $I(V)$ curves in atomic nanowires of pure superconducting metals \cite{Scheer1998Nature}. However, if the central constriction region contains materials which become superconducting only via the proximity effect, a considerable energy smearing is observed on the $I(V)$ characteristics. This phenomenon is best characterized by a superconducting tunneling junction where the $\Gamma$ width of the peaks at $eV\approx\pm2\Delta$ of the $dI/dV$ curves reflects this energy smearing \cite{Torok2020,Proximity-Scheer}. According to our previous work on the subgap $I(V)$ curves of Nb$_2$O$_5$ junctions the active volume of transition metal-oxide RS systems contains non-superconducting suboxides, and therefore the proximity-effect-induced energy smearing is a significant phenomenon which restricts the energy resolution of the measurements \cite{Torok2020}. Figure~\ref{Fig3}b shows $dI/dV$ characteristics of ScS tunnel junctions in pure Ta atomic wires (green) and in tantalum-oxide RS devices (black), yielding a broadening value of $\Gamma_{\rm ox.}\approx 600~\mu$V for the latter. During our analysis the simulated $I(V)$ curves are convoluted with a Gaussian function simulating this energy broadening. 

\newpage
\section{Cutoff dependence of the simulated $I(V)$ characteristics and histograms.}

\begin{figure}[h!]
		\includegraphics[width=0.7\linewidth]{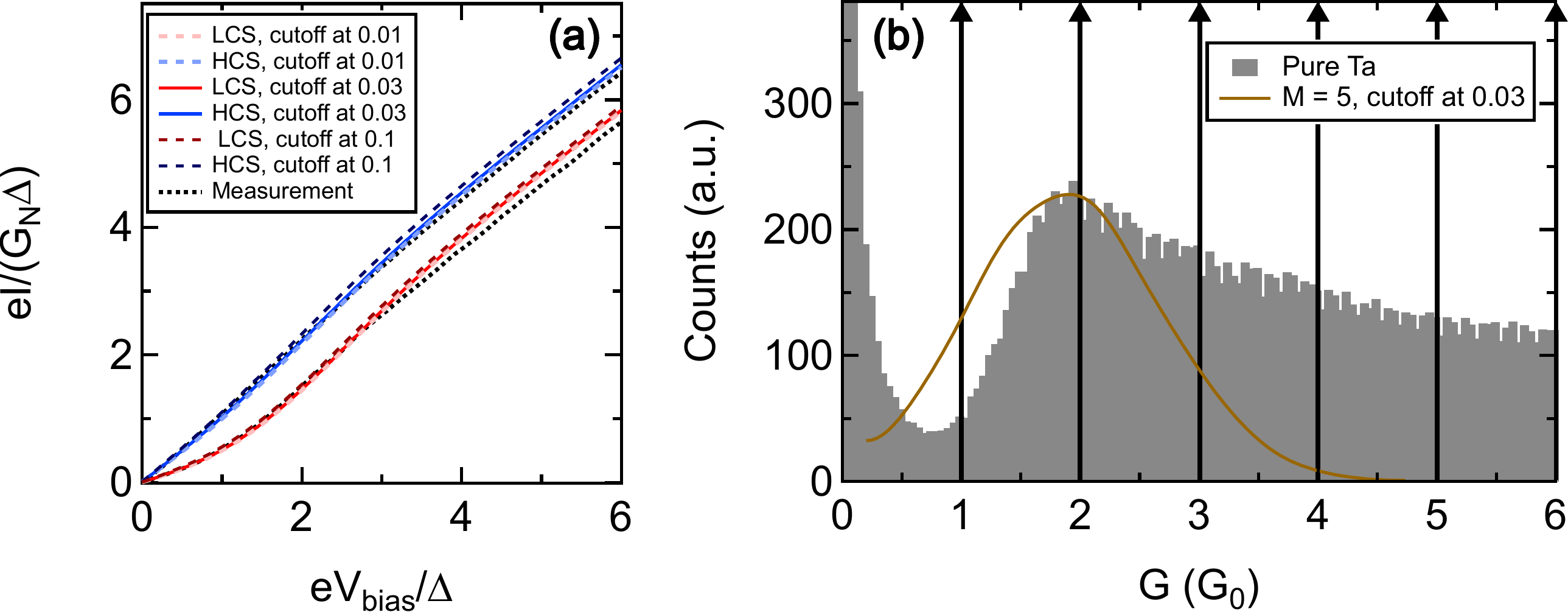}
		\caption{\textbf{Role of the transmission density cutoff in the simulated $I(V)$ curves and histograms.} (a) The measured ScS $I(V)$ curves (black dots) and the simulated ScS $I(V)$ curves (blue and red line, $\tau=0.03$ cutoff) from Fig.~3c of the manuscript are compared to the simulated $I(V)$ curves generated by the same transmission density (same $\alpha$ value), but with different cutoff. The dark blue \& dark red / light blue \& light red dashed lines respectively correspond to $\tau=0.1$ and $\tau=0.01$ cutoff. (b) Simulated conductance histogram using the $\rho^\mathrm{RMT}(\tau)$ transmission density, $M=5$ conductance channels and $\tau=0.03$ cutoff (brown line). As a reference, the conductance histogram of pure Ta atomic wires is reproduced from Fig.~1d of the manuscript (grey histogram). The black Dirac delta peaks illustrate the extreme case of prefectly transmitting open channels, which would yield quantized conductance values in integer units of $G_{0}$.}
		\label{Fig4}
\end{figure}

According to the theory of multiple Andreev reflections \cite{Cuevas1996}, both the leading subgap current contribution and the excess current \cite{Senkpiel2022} are quadratically suppressed for a poorly transmitting channel, $\tau \ll 1$. Accordingly, the shape of the simulated ScS $I(V)$ characteristics are strongly susceptible/insensitive to the weight of the highly/poorly  transmitting channels in the transmission density. This also means, that the precise value of the low-transmisson cutoff of the transmission density function does not have a significant impact on the simulated $I(V)$ curves. In Fig.~\ref{Fig4}a this is demonstrated for various cutoff values: $\tau=0.01$ and $\tau=0.03$ yield hardly distinguishable $I(V)$ characteristics, and a minor deviation (i.e. a tiny increase of the excess current) is observed at   an even larger cutoff of $\tau=0.1$.

Whereas the simulated ScS $I(V)$ curves are insensitive to the bottom cutoff of the density function, the average transmission and the channel number at a certain conductance do depend on it.  We have chosen the physically reasonable cutoff value of $\tau=0.03$ such, that at this cutoff the peak position of the simulated histogram with $\rho^\mathrm{RMT}(\tau)$ transmission density and $M=5$ channels (brown line in Fig.~\ref{Fig4}b) coincides with the central conductance of the experimental histogram peak, $G \approx 1.9\,$G$_0$. The latter demonstrates the most probable conductance of a single-atom Ta junction, for which $\approx 5$ open conductance channels are expected according to theoretical considerations \cite{Cuevas_PRL_1998}. As a comparison, a cutoff of $\tau=0.01$ ($\tau=0.1$) would respectively yield $M=8$ ($M=4$) open channels at the most probable single-atom conductance.

The simulated histogram with $\rho^\mathrm{RMT}(\tau)$ density and 5 channels (brown curve in Fig.~\ref{Fig4}b) yields a rather broad conductance distribution, lacking any conductance quantization features. This means, that the presence of a few highly open conductance channels in $\rho^\mathrm{RMT}(\tau)$ are sensitively detected by the ScS subgap spectra, but the conductance histrogram rather reflects the overall broad variation of the transmission values, which agrees with the very broad, non-quantized distribution of the experimental histograms for transition metals. For any sign of conductance quantization, a transmission density with much stronger weight at the high transmission end, $\tau\approx 1$ would be required, which is only possible in metals, where solely the highly delocalized $s$-electrons dominate the transport \cite{Scheer-Cuevas}. The black Dirac-delta peaks in Fig.~\ref{Fig4}b illustrate the most transmissive limit, where all the open channels are fully open ($\tau=1$). This transmission density (black curve in Fig.~2b of the manuscript) would oviously yield precisely quantized conductance values. Such \emph{clear} conductance quantization, however, is merely expected in adiabatic semiconductor quantum point contacts. 


\newpage

\bibliographystyle{achemso}
\bibliography{References}